\documentclass[aps,prx,twocolumn,superscriptaddress,longbibliography,nofootinbib]{revtex4-2}
\usepackage{amsmath,amssymb}
\usepackage{graphicx}
\usepackage{xcolor}
\usepackage{booktabs}
\usepackage{verbatim}
\usepackage{multirow}
\usepackage{siunitx}
\usepackage[ruled,linesnumbered,noend]{algorithm2e} 
\SetAlFnt{\small}\SetAlCapFnt{\small}\SetAlCapNameFnt{\small}

\usepackage[colorlinks=true,allcolors=blue]{hyperref}

\usepackage[normalem]{ulem}
\usepackage{listings}
\definecolor{pykw}{HTML}{A8501E}\definecolor{pyapi}{HTML}{004488}\definecolor{pystr}{HTML}{117733}\definecolor{pycom}{HTML}{6E6E6E}
\lstdefinestyle{pybase}{language=Python, keywordstyle=\color{pykw}\bfseries, stringstyle=\color{pystr}, commentstyle=\color{pycom}\itshape, emph={CapacitancePredictor,MLSimulator,PlanarCapacitanceDB,Analyzer,load_model,predict,predict_capacitance,to_sim_results,designs,labels,embedding}, emphstyle=\color{pyapi}, showstringspaces=false, upquote=true, columns=fullflexible, keepspaces=true}
\lstdefinestyle{py}{style=pybase, basicstyle=\ttfamily\footnotesize, breaklines=true, breakatwhitespace=true, frame=lines, framesep=3pt, framerule=0.4pt, rulecolor=\color{pycom}, xleftmargin=0pt, xrightmargin=0pt, aboveskip=6pt, belowskip=4pt, linewidth=\columnwidth}
\lstdefinestyle{pyinline}{style=pybase, basicstyle=\ttfamily}
\newcommand{\CNS}{\ensuremath{C_{\mathrm{NS}}}}
\newcommand{\epseff}{\ensuremath{\varepsilon_{\mathrm{eff}}}}
\newcommand{\fF}{\ensuremath{\,\mathrm{fF}}}
\newcommand{\MHz}{\ensuremath{\,\mathrm{MHz}}}
\newcommand{\um}{\ensuremath{\,\mu\mathrm{m}}}
\newcommand{\CR}{\ensuremath{C^{\mathrm R}}}
\newcommand{\ppg}{panels per smallest gap}
\newcommand{\AtlFinN}{348}

\begin{document}

\title{QCap: a physics-corrected graph neural network for the capacitance matrix of any planar superconducting device layout}

\author{Sadman Ahmed Shanto}
\email{shanto@usc.edu}
\affiliation{Department of Physics and Astronomy, University of Southern California, Los Angeles, California 90089, USA}
\affiliation{Center for Quantum Information Science and Technology, University of Southern California, Los Angeles, California 90089, USA}

\author{Saikat Das}
\email{saikatda@usc.edu}
\affiliation{Department of Physics and Astronomy, University of Southern California, Los Angeles, California 90089, USA}
\affiliation{Center for Quantum Information Science and Technology, University of Southern California, Los Angeles, California 90089, USA}

\author{Eli M. Levenson-Falk}
\email{elevenso@usc.edu}
\affiliation{Department of Physics and Astronomy, University of Southern California, Los Angeles, California 90089, USA}
\affiliation{Center for Quantum Information Science and Technology, University of Southern California, Los Angeles, California 90089, USA}
\affiliation{Ming Hsieh Department of Electrical \& Computer Engineering, University of Southern California, Los Angeles, California 90089, USA}
\affiliation{Quantum Elements, Inc., Westlake Village, California 91361, USA}

\begin{abstract}
Electromagnetic simulations of device layouts to extract capacitance are computationally intensive, a major obstacle to accurate design of circuit devices. Machine learning models can bypass these simulations, but previous approaches have been restricted to layouts using only particular components, %
and require thousands of simulations of those same components as training data. We present QCap, a model that accurately predicts the Maxwell capacitance matrix of an \textit{arbitrary} superconducting device layout in seconds, compared to minutes to hours of Ansys Q3D simulations. We develop a novel design-embedding algorithm that converts the layout into a universal graph format, allowing the model to work with any layout created with any tool. To achieve high accuracy, we obtain a coarse physics prior from a fast method of moments simulation of capacitance, then train the model to correct this prior. On 4,009 test designs withheld from training data the model outputs agree with Q3D simulations to a median error of 0.177\%, and on never-seen component families it stays below 1\% error. We demonstrate the model's scaling by obtaining the capacitance matrix of a 9-superconducting-qubit device in 12 seconds on a laptop (compared to 19.9\,h and 228\,GB of memory in Ansys) with sub-percent error, and capacitance of a (too large to simulate) 100-qubit, 781-conductor device in 3.4\,min on a laptop. We provide the open-source models and data on SQuADDS and Hugging Face. Our results provide a powerful tool for large-scale accurate circuit design and guide the development of models to bypass other physical simulations.

\end{abstract}

\maketitle

\section{Introduction}
\label{sec:intro}

Electromagnetic (EM) simulations to predict analog circuit device behavior are essential for accurate design. As an example, we focus on superconducting quantum circuits, which are among the most mature platforms for fault-tolerant quantum computing. They have been operated below the threshold of the surface code~\cite{Fowler2012,GoogleQEC2025} and are fabricated with lithographic processes that scale to large chips~\cite{Krantz2019,Kjaergaard2020,Blais2021}. Progress on this platform runs in learning cycles of design, fabrication and measurement~\cite{LevensonFalk2025Review}. Faster, more accurate predictions of device behavior can make these cycles shorter and more effective, allowing faster progress towards utility-scale quantum processors.

\begin{figure*}[!t]
\centering
\includegraphics[width=\textwidth]{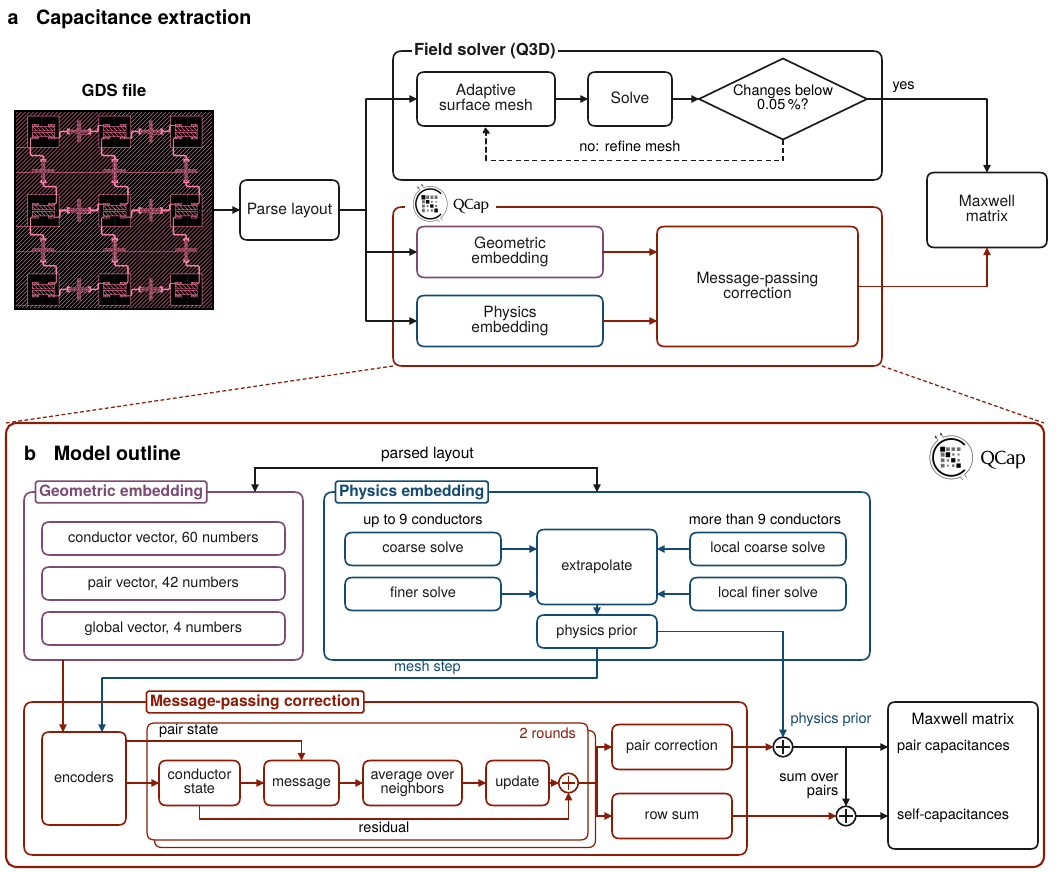}
\caption{Workflow for capacitance extraction from a layout, by using a traditional field solver (top) and by using QCap, the model presented in this work. (a) The input is a GDS file, here the nine-qubit grid of Sec.~\ref{sec:res-windows}. A solver such as Q3D meshes the layout, solves it, and refines the mesh until the convergence criterion is met. QCap instead embeds the geometry and a physics prior and corrects the prior with a message-passing network. (b) A detailed depiction of QCap. %
The geometric embedding gives a vector for every conductor, for every pair of conductors and for the whole layout. The physics embedding solves the layout on a coarse and a finer mesh, in local windows when it has more than nine conductors, and extrapolates the two solves to the physics prior. The message-passing correction updates the state of every conductor in two rounds. Its two heads then give a correction to the prior for every pair, within a correction limit, and a positive row sum for every conductor, which together give the pair capacitances and the self-capacitances of the Maxwell matrix.} %
\label{fig:overview}
\end{figure*}

\begin{figure*}[!t]
\centering
\includegraphics[width=\textwidth]{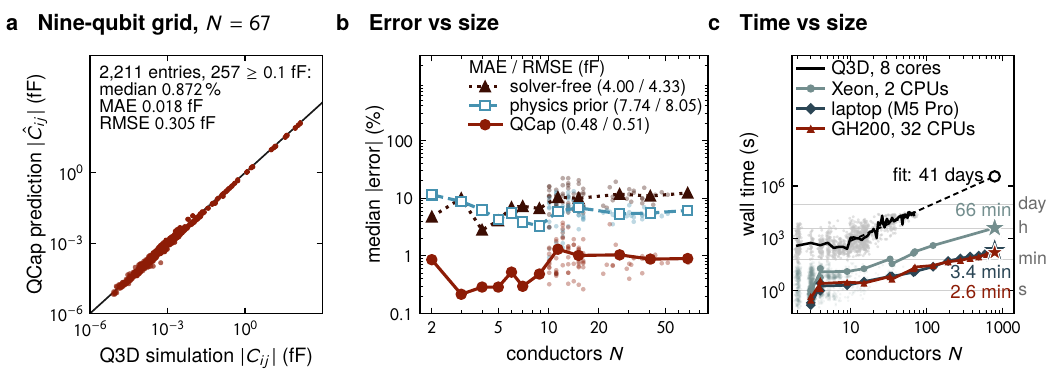}
\caption{(a) Agreement between QCap and Q3D simulations for capacitances of the nine-qubit grid device (Sec.~\ref{sec:res-windows}). (b) Layout-mean median error against the number of conductors $N$ on all test datasets of Table~\ref{tab:cohorts}. The legend gives the MAE and RMSE of each predictor over these layouts. (c) Wall time against $N$ of QCap on different hardware and of Q3D per design on eight cores. Q3D takes 6.8\,h in median at $N=67$-68, and a power-law fit to its 595 designs with $N\ge10$, $t\propto N^{1.95}$, extrapolates to 41 days for the 100-qubit chip ($N=781$), which QCap predicts in 2.6\,min to 66\,min depending on the hardware.}
\label{fig:overview-results}
\end{figure*}

EM simulations of capacitance are particularly important for superconducting device design. The shunt capacitance of a transmon, the most common superconducting qubit, sets its charging energy $E_C=e^2/2C_\Sigma$, and with it the anharmonicity $\eta\approx-E_C$ and the transition frequency $\omega_q\approx\sqrt{8E_JE_C}-E_C$~\cite{Koch2007} (where $E_J$ is the energy of a Josephson element). Capacitances couple qubits to readout resonators, buses, drive lines and tunable couplers, and every parasitic capacitance causes unwanted interactions~\cite{Krantz2019,Forati2026}. The capacitances of an $N$-conductor device are described in its $N\times N$ Maxwell capacitance matrix $C_{ij}$~\cite{Blais2021,Minev2021LOM}. Typically capacitance is computed with a quasi-static field solver such as Ansys Q3D~\cite{AnsysQ3D} or Palace~\cite{Palace,Sommers2025SQDMetal}. Open layout tools~\cite{QiskitMetal,KQCircuits,gdswell} and design databases~\cite{Shanto2024SQuADDS} make candidate layouts cheap to draw, and quantization methods turn a capacitance matrix or an eigenmode solution into a circuit Hamiltonian~\cite{Minev2021LOM,scqubits,Chitta2022,Minev2021EPR}, including for circuits with distributed resonators~\cite{Kunz2026}. Repeating capacitance simulations as a design is iteratively refined can be extremely computationally costly. Each simulation takes minutes for a two-conductor capacitor, 1-2 hours for a qubit with several coupling capacitors, about half a day for a qubit embedded with four tunable couplers in a grid (Fig.~\ref{fig:overview-results}(c)), and becomes infeasible for large devices.

Several approaches are possible to bypass these computationally intensive simulations, but each has its limitations. Physical models such as conformal-mapping formulas for coplanar structures~\cite{Wen1969,Gevorgian1996,Simons2001} assume a closed form for the geometry at hand, reducing accuracy. %
Fast integral-equation solvers and the multipole and hierarchical-matrix methods that speed them up~\cite{Nabors1991FastCap,Greengard1987,Hackbusch1999,Bebendorf2000} still mesh and solve every layout from scratch, while a mesh-free random-walk solver guided by a network~\cite{Rodriguez2026DeepRWCap} still solves each layout anew. Machine-learned surrogates~\cite{MLQubitDesign,NugrahaShao2023,Yaker2026} map between the design parameters of one component and its Hamiltonian parameters, forward or inverse. However, their input is the parameter string of a particular layout generation tool, so such a model cannot be evaluated on another component, on a layout drawn by another tool, or on a chip with many compponents, and each new component requires thousands of new training simulations. Graph neural networks have been trained to predict capacitance of a conventional integrated circuit directly from a layout~\cite{Liu2024GNNCap, Huang2026AttentionCap, Rodriguez2026CapBench}, for predicting the parasitics from a circuit schematic before the layout is drawn~\cite{Ren2020ParaGraph}, and to choose the gate frequencies of large superconducting processors~\cite{Ai2025GNN}. In every case the network maps its input straight to the answer, so all the physics must be reflected in and learned from the training data, requiring large training data and limiting accuracy on prediction of never-before-seen devices. %
In contrast, training a model once and reusing it across tasks~\cite{Pan2010,Devlin2019,Bommasani2021} can be far more accurate and efficient, but has not yet been demonstrated for EM simulation.

In this paper we present QCap, a machine-learned model that computes the full Maxwell capacitance matrix of \textit{any} planar layout with \textit{any} number of conductors \textit{from the layout .gds file alone}. Our model matches EM solver results in seconds per component. %
The model is based on two advancements. First, we develop a geometric embedding of the layout, a \textit{richer representation than a parameter string} that uses a small number of parameters to capture the geometry of an arbitrary device in a graph whose nodes are the conductors and whose edges are conductor pairs. 
The same functions act on every node and every edge, so the same network can serve any layout with any number of conductors from any drawing tool, including components it has never seen. %
The second advance is our use of a \emph{cheap physics prior} to quickly get a rough physical prediction of capacitance. The neural network then  %
learns only a bounded residual on the prior. %
Learning a map that fully encodes Maxwell's equations would be a hard training problem, but the prior turns it into a small local correction, making the model accurate enough to substitute for the field solver %
despite limited training data (about 35,000 Q3D solutions). The model gives capacitance predictions on in-family test designs that differ from Q3D solves  by a median relative error of 0.177\%, a median absolute error (MAE) of 0.016\fF, and a root mean square error (RMSE) of 0.025 \fF, %
and are within 0.408\% on a whole component family held out of training (MAE 0.0413\fF, RMSE 0.0591\fF). %
We show that the model can achieve percent-level accuracy on large devices and can predict device capacitances far beyond the scale that Q3D can handle. Despite all training data having the same substrate thickness and dielectric constant, the model maintains percent-level accuracy predicting other thicknesses and dielectrics when a small correction factor is incorporated. %

In Section~\ref{sec:embedding} we explain the embedding protocol.  Sec.~\ref{sec:learning} introduces QCap as well as two variants (Fig.~\ref{fig:overview-results}(b)) and their training.  Sec.~\ref{sec:results} contains the main results. Details of the methods and secondary results are in the appendices.%
\section{Embedding protocol}
\label{sec:embedding}
Consider a family of components $F$ with design space parameters $\boldsymbol\theta_F$ --- for instance, two rectangular capacitor pads with parameters length, width, height, and spacing. 
A surrogate trained on $\boldsymbol\theta_F$ of one component family $F$ learns a map that cannot be evaluated on a second family, or even on the same family parametrized in a different way (i.e., made with a different design tool). What current ML approaches to chip design lack is an \emph{embedding}, a map $\Phi$ that takes %
layouts (objects with any number of conductors, drawn with any tool) into a space a learning algorithm can operate on~\cite{Mikolov2013,Zaheer2017,Bronstein2021}. %
Learning a complicated physics map from scratch may require large amounts of training data, and a good embedding needs to be able to work at any scale. We therefore require $\Phi$ to be universal (defined for any planar layout independent of the design tool used), geometric (describing each conductor and each coupling in physical units), local and size-invariant (a qubit inside a grid looks like the same qubit on its own chip), physical (carrying a physics block from which the network learns), and fast at every system size.

The eventual goal is to replace the results of capacitance simulations. To train, validate, and test our approach, we use a data set of Ansys Q3D simulations of various components. Simulation details are given in Appendix~\ref{app:q3d}. The simulations' own noise, from the convergence of the pass sequence and the mirror asymmetry of symmetric designs, is below 0.02\% (0.005\fF). An independent finite-element solver, Palace~\cite{Palace}, on meshes from meshwell~\cite{Meshwell}, agrees with Q3D to within 1.7\% in median over the mutual entries (MAE at most 2.2\fF, RMSE at most 1.9\fF) on the one-, five- and nine-qubit grids and on 14 component designs of six families. Overall, we have simulated results on 39,962 layouts that we split into 35,354 for training and 4,608 for validation, and a further 5,902 layouts form the test sets (Table~\ref{tab:data}). Figure~\ref{fig:dataset-atlas} shows some representative layouts from our dataset.

\begin{table}[!t]
\caption{The dataset in five groups, with the splits for training, validation, and testing, the number of conductors $N$ including ground, and representative layouts. Figure~\ref{fig:dataset-atlas} shows layouts of every family in each group, and Table~\ref{tab:data-full} of Appendix~\ref{app:data} lists every family and test set.}
\label{tab:data}
\scriptsize\setlength{\tabcolsep}{2pt}
\begin{ruledtabular}
\providecommand{\grp}[2]{\begin{tabular}{@{}l@{}}#1\\#2\end{tabular}}
\begin{tabular}{@{}lccccc@{}}
Group & $N$ & Training & Validation & Test & Rep. Layouts\\
\hline
\grp{Capacitor}{families} & 3--7 & 32{,}073 & 4{,}071 & 4{,}009 & {\raisebox{-0.5\height}{\includegraphics[width=1.09cm,height=0.76cm]{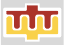}}}\,{\raisebox{-0.5\height}{\includegraphics[width=1.09cm,height=0.76cm]{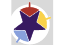}}}\\[1pt]
\grp{Qubit}{families} & 3--5 & 735 & 93 & 926 & {\raisebox{-0.5\height}{\includegraphics[width=1.09cm,height=0.76cm]{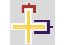}}}\,{\raisebox{-0.5\height}{\includegraphics[width=1.09cm,height=0.76cm]{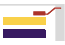}}}\\[1pt]
\grp{Generated}{layouts} & 2--68 & 2{,}546 & 295 & 364 & {\raisebox{-0.5\height}{\includegraphics[width=1.09cm,height=0.76cm]{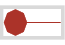}}}\,{\raisebox{-0.5\height}{\includegraphics[width=1.09cm,height=0.76cm]{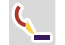}}}\\[1pt]
\grp{Held-out}{sets} & 2--18 & 0 & 149 & 600 & {\raisebox{-0.5\height}{\includegraphics[width=1.09cm,height=0.76cm]{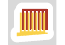}}}\,{\raisebox{-0.5\height}{\includegraphics[width=1.09cm,height=0.76cm]{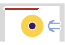}}}\\[1pt]
\grp{Multi-qubit}{grids} & 15--67 & 0 & 0 & 3 & {\raisebox{-0.5\height}{\includegraphics[width=1.09cm,height=0.76cm]{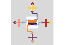}}}\,{\raisebox{-0.5\height}{\includegraphics[width=1.09cm,height=0.76cm]{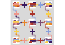}}}\\
\hline
Total & & 35{,}354 & 4{,}608 & 5{,}902 & \\
\end{tabular}
\end{ruledtabular}
\end{table}

\subsection{Bipartite embedding structure}
\label{sec:emb-two}

\begin{figure*}[!tb]
\centering
\includegraphics[width=\textwidth]{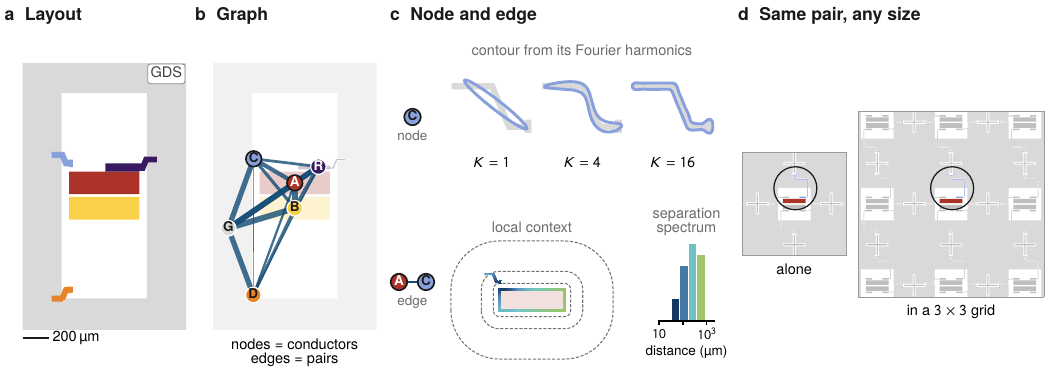}
\caption{Walkthrough of the geometric embedding. (a) A TransmonPocket read from its .gds file as six conductors: the ground (gray), the two pads A (red) and B (yellow) and the three couplers C, R and D, drawn thicker than in the actual design for visibility. (b) The same layout as a graph, with a node for every conductor and an edge for every pair; edge width indicates the physics prior $\log c^{\mathrm R}_{ij}$. (c) %
The 60 entries of a node vector are 11 shape scalars, 33 Fourier magnitudes of the outer contour, and 16 width spectra. The top row shows coupler C rebuilt from $K=1$, 4 and 16 of its Fourier harmonics. The 42 entries of a pair vector are the separation spectrum with the minimum gap, three coupling moments, a conformal coupling, two shielding shares, a ground flag, and the local context. The bottom row shows for the pair of A and C the local context within 50, 150 and 400\um\ (dashed), with each boundary sample colored by its distance to the other conductor. On bottom right is the separation spectrum, how much boundary of one conductor faces the other at each distance. (d) A pad and its bus line (circled) have the same node local vectors alone on a chip and inside a $3\times3$ grid of 67 conductors, and, once standardized, the same pair vectors except for the two shielding shares. The whole embedding takes under 0.1\,s on one laptop core.}
\label{fig:walkthrough}
\end{figure*}

The embedding has a geometric part and a physics part,
\begin{equation}
\Phi(\mathcal L)=\Big(\underbrace{\{\mathbf n_i\}_{i<N},\ \{\mathbf e_{ij}\}_{i<j<N},\ \mathbf u}_{\text{geometric}},\
\underbrace{\CR,\ \{s_{ij}\}}_{\text{physics}}\Big),
\label{eq:embedding}
\end{equation}
i.e., a set of fixed-length vectors on a graph whose nodes $\{\mathbf{n_i}\}$ are the $N$ conductors of the layout and whose edges $\{\mathbf{e_{ij}}\}$ are their pairs, together with some global information $\mathbf{u}$ and the outputs of a cheap physics solve $C^\mathrm{R}$ and $\{s_{ij}\}$ (defined in Sec.~\ref{sec:emb-prior}). Note that a single vector of fixed length could not do this job, because the number of conductors varies from layout to layout and the matrix to be predicted has $N^2$ entries. Figure~\ref{fig:walkthrough} follows the construction of the geometric part on one layout.

\subsection{Geometric embedding}
\label{sec:emb-geometric}

The idea of the geometric embedding algorithm is to provide the network maximum geometric information in a form that means the same thing for every component. From the input .gds file we unite all polygons of the ground layer into one ground net and take each galvanically connected body of the signal layer as one conductor, so $N$ is read directly from the file (Fig.~\ref{fig:walkthrough}(a)). Every pair of conductors, the ground included, is joined by an edge, forming a complete graph with $N(N-1)/2$ edges. A network fed this graph (Fig.~\ref{fig:walkthrough}(b)) applies the same functions to every node and edge, so its parameter count does not depend on $N$ and relabeling the conductors only permutes the output (Sec.~\ref{sec:ml-network}).

The conductor vector $\mathbf n_i\in\mathbb R^{60}$ describes the shape of each conductor, ground included, by shape scalars, the Fourier magnitudes of its outer contour~\cite{Granlund1972}, and a width spectrum. An edge holds the pair vector $\mathbf e_{ij}\in\mathbb R^{42}$, which says how much metal of $i$ faces how much metal of $j$, at what distance, and what surrounds them (Fig.~\ref{fig:walkthrough}(c)). It holds a \emph{separation spectrum} of the facing boundary, i.e., how much boundary faces the other conductor in each of ten distance bins, \emph{coupling moments} that weight this spectrum as a coupling would, a \emph{conformal coupling} from the coplanar-strip formula~\cite{Simons2001}, two \emph{shielding shares}, which are the fraction of each conductor's total coupling that this pair carries, and the \emph{local context}: the fractions of ground, other metal and etched substrate within 50, 150 and 400\um. Importantly, almost nothing in a pair vector refers to the chip size, the pocket size or the total number of conductors. Only the two shielding shares, normalized by each conductor's total coupling, see the rest of the chip, and in a reference device they move by 0.1 interquartile range between a qubit alone on its chip and the same qubit inside a $3\times3$ grid of 67 conductors, with every other entry identical (Fig.~\ref{fig:walkthrough}(d)). This is what lets a correction learned on components act on a chip. Lastly, a global vector $\mathbf u\in\mathbb R^4$ holds the logarithms of the chip area, the pocket area, the substrate thickness and the effective permittivity. Performing the embedding takes $\sim 0.1$ s for a component and a few seconds for a 68-conductor layout (see Appendex~\ref{app:embedding} for a detailed overview).

An ML system can be trained on this embedding to learn any layout-level target (e.g. capacitance), and can generalize to layouts outside its training families. %
In practice, however, its effectiveness depends strongly on the diversity and size of the training data (Sec.~\ref{sec:res-sota}, Appendix~\ref{app:res-scaling}). Unfortunately, large data sets of superconducting device EM simulations are not available. %
We therefore use a physics prior to reduce training requirements.

\subsection{Fast physics prior}
\label{sec:emb-prior}

Learning the physics of electrostatics with a network is a challenging task. It is far easier for a network to correct an inaccurate physics result to match the reference Q3D result. %
This lets the model learn from little data and act on components it has never seen, because the correction is only a function of local geometry that reads the same on every layout. Importantly, the prior is also where we can put what we know about the physics of the problem, and the network learns the rest. Physics that the data cannot teach enters the model here, and it is here that a physical expectation can be written in to steer the model (an example of this is how we calibrate for the substrate properties as explained in Sec.~\ref{sec:res-stack}). While we use this approach to train models for predicting capacitance matrices, \textit{it is a completely general physics-informed learning method} that can be adapted for any use case where we want to train models but where reference data is scarce.

Let the \textit{physics prior} $C^{\mathrm R}$ be the Maxwell capacitance matrix of a cheap but physically complete solve of the same layout (described below) and $c^\ast$ be a reference training capacitance matrix. The off-diagonal elements are the negatives of the capacitances between conductors, and the on-diagonal elements are the total capacitance of a conductor to all other conductors and ground. Entry by entry, we define the residual $\varepsilon_{ij}$ as the difference between the logarithms of the prior and the reference,
\begin{equation}
\log c^{\ast}_{ij}=\log C^{\mathrm R}_{ij}+\varepsilon_{ij},
\label{eq:residual}
\end{equation}
The network learns this residual $\varepsilon_{ij}$, as in $\Delta$-learning and multifidelity modeling~\cite{Ramakrishnan2015Delta,Kennedy2000}. %
The prior need not be accurate, it need only encode the basic physics so that the network's correction is bounded. This bound parameter needs to be measured from the dataset and depends on how the prior was defined. The network of Sec.~\ref{sec:learning} therefore predicts $\varepsilon_{ij}$ only within that bound, so where the data runs out the prediction stays close to the physics. The residual is also the same function for every component that presents the same local situations, which is why a correction learned on three-conductor capacitors can act on a many-conductor chip (Table~\ref{tab:grid-without}).

The physics prior we use in our work has been carefully chosen to optimize for fast computation speed with just enough accuracy to make training our networks effective. We use a method of moments on the metal alone~\cite{Harrington1968,Nabors1991FastCap} (Appendix~\ref{app:m-prior}) where the conductors are treated as sheets of zero thickness on a substrate of permittivity $\varepsilon_r$ and thickness $t$, with vacuum above and below it. In contrast, the reference Ansys Q3D solve meshes the entire chip, its substrate, and the box around it and refines that mesh pass after pass until the matrix converges. %
To compute the prior fast we only discretize the metal %
coarsely and account for the substrate through a Green's function kernel.

We treat the substrate as an image series in the substrate thickness $t$ and dielectric constant $\varepsilon_r$~\cite{Sometani2000}, as the closed-form Green's functions of layered media do in microwave and capacitance extraction~\cite{Chow1991,Oh1994} (Eq.~\eqref{eq:substrate}). This allows us to reduce the problem to a uniform medium of effective permittivity $\epseff=(\varepsilon_r+1)/2$ for distances well below $t$ and to the vacuum kernel far beyond it, because the field lines between distant conductors mostly leave the substrate. This series also makes the substrate thickness and permittivity explicit inputs of the prior, allowing the model to work on unseen substrates %
(see Section~\ref{sec:res-stack}). The charge on the metal is expanded in flat panels, Eq.~\eqref{eq:mom}, which makes the solving very fast: a fraction of a second for a component (0.075\,s in median on two CPUs of an EPYC node, Table~\ref{tab:sota}) compared to the minutes to hours required when accurately solved via Q3D (Table~\ref{tab:cohorts}).

To maximize the accuracy of our prior, we make use of the fact that the error of a coarse solve is mainly set by its mesh resolution. %
The charge on a thin film piles up at its edges as $d^{-1/2}$ with the distance $d$ to the edge. Flat panels of size $h$ cannot follow that, so every solve misses some edge charge, and the error is first order in $h$. In our data the error is always the same sign, with the prior coming out too low (Appendix~\ref{app:firstorder}). Luckily, one Richardson step~\cite{Richardson1911} removes this first-order error: 

From two solves at panel sizes $2h$ and $h$, we expect the coarser solve to make roughly twice the error of the finer one. We thus project the refined answer as
\begin{equation}
\log C^{\mathrm R}_{ij}=2\log C^{h}_{ij}-\log C^{2h}_{ij},
\label{eq:richardson}
\end{equation}
applied entry by entry in $\log|C|$. %
On the validation layout dataset, we find that this halves the prior's error for the finer solve alone and at less than twice the cost of that solve (Appendix~\ref{app:richardson}). This optimized prior, $\CR$, is what we use for the physics embedding. The mesh step $s_{ij} \equiv \log C^{h}_{ij}-\log C^{2h}_{ij}$ is kept with every entry as the one physics-derived input of the pair vector.

This prior's compute time is proportional to the size of the chip ($\mathcal{O} (\mathrm{area}^2)$ in memory and $\mathcal{O}(\mathrm{area}^3)$ in time). We therefore make one further optimization to speed things up even more for larger devices. For layouts of more than nine conductors we solve the prior in \emph{windows} (Fig.~\ref{fig:windows}(a)). %
Groups of at most eight conductors are excited one at a time inside a window that leaves nearby features unchanged and merges everything farther away into the ground, so the size of a window is set by the metal around its core and not by the chip (Appendix~\ref{app:windows}). For pairs far apart that cannot be contained in the same window, we use the %
closed form of a far-field coupling through a grounded plane~\cite{Martinis2014Coupling},
\begin{equation}
c^{\mathrm{far}}_{ij}=\frac{\epseff\,\varepsilon_0\,\tilde A_i\tilde A_j}{\pi\,d_{\mathrm{eff}}^{3}},\qquad
d_{\mathrm{eff}}^2=d^2+\frac{\tilde A_i+\tilde A_j}{\pi},
\label{eq:far}
\end{equation}
i.e., the field of a dipole through an opening of effective area $\tilde A_i$ collected by a second opening at distance $d$. The effective area $\tilde A_i=A_i+\alpha\mathcal P_i\bar g_i$ is the area $A_i$ of conductor $i$ plus a share $\alpha$ of its perimeter $\mathcal P_i$ times its mean gap $\bar g_i$ to the nearest other metal, i.e., the conductor together with part of the gap around it, with $\alpha=0.497$ fitted once on training layouts. Eq.~\eqref{eq:far} as written holds for two conductors in the same ground pocket. Across ground pockets the field below the plane sees a substrate of finite thickness, and $\epseff$ is replaced by a distance-dependent $\varepsilon_{\mathrm{far}}(d)$ as shown in Fig.~\ref{fig:windows}(b) (Appendix~\ref{app:fp-far}). The cost of the optimized prior now grows with the number of conductors, about $N^{1.4}$ on the grids of Sec.~\ref{sec:res-windows}, and not with the chip area. %

\section{The learning problem}
\label{sec:learning}

The input of QCap is a .gds layout file and its output is the full Maxwell matrix $\hat C$ of its $N$ conductors, with $N$ read from the layout file. Section~\ref{sec:embedding} showed how we go from the layout to its embedding and in this section we explain how we go from the embedding to the output of the capacitance matrix (illustrated in Fig.~\ref{fig:overview}(b)).

\subsection{Model architecture}
\label{sec:ml-network}

\begin{figure}[!tbp]
\centering
\includegraphics[width=\columnwidth]{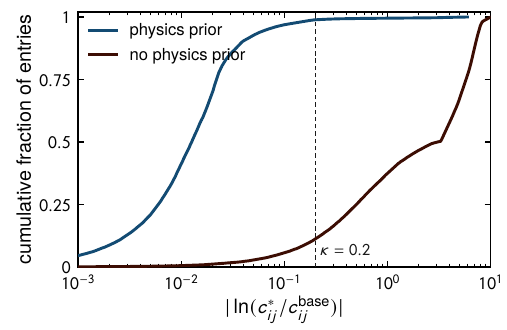}
\caption{The cumulative distribution of $|\ln(c^{\ast}_{ij}/c^{\mathrm{base}}_{ij})|$ (the correction the network must learn) over the mutual entries of at least 0.1\fF\ of the validation layouts. %
Here $c^{\mathrm{base}}_{ij}$ is the physics prior $c^{\mathrm R}_{ij}$ (Eq.~\eqref{eq:richardson}), or with no physics prior the training median of the pair kind (signal-signal or signal-ground). The choice of prior determines the bound $\kappa$ of the correction.}
\label{fig:physics-concept}
\end{figure}

Let $h_i$ be the encoding of conductor $i$ and $r_{ij}$ the symmetric output of the pair head of the network (described below). QCap predicts
\begin{equation}
\begin{aligned}
\log\hat c_{ij}&=\log C^{\mathrm R}_{ij}+\kappa\tanh r_{ij},\\
\hat b_i&=\exp\!\big(\mu_{\mathrm{role}(i)}+q(h_i)\big),\\
\hat C_{ii}&=\sum_{j\neq i}\hat c_{ij}+\hat b_i,\qquad \hat C_{ij}=-\hat c_{ij}\ (i\neq j),
\end{aligned}
\label{eq:prediction}
\end{equation}
Here $q$ is a node head and $\mu_{\mathrm{role}}$ is the median over all training data conductors of the same role (ground or signal) of the log of the sum of the corresponding matrix row. %
The residual $\varepsilon_{ij}$ of Eq.~\eqref{eq:residual} is parametrized as $\kappa\tanh r_{ij}$, so every mutual entry lies within a factor $e^{\pm\kappa}$ of the extrapolated physics, and the error of a prediction can never exceed the error of the physics by more than $\kappa$,
\begin{equation}
|\log\hat c_{ij}-\log c^{\ast}_{ij}|\le|\log c^{\mathrm R}_{ij}-\log c^{\ast}_{ij}|+\kappa .
\label{eq:bound}
\end{equation}
The parameter $\kappa$ is the size of the correction the network is allowed. From our datasets, we find that the optimized prior is off by up to about 20\% on all but a few training entries, so we set $\kappa=0.2$, a factor of $e^{\pm0.2}$  (Appendix~\ref{app:kappa}), which allows our model to accurately correct every entry whose prior is already that close to the reference. %
The mutual entries are positive and the matrix is symmetric because $r_{ij}=r_{ji}$, and each row sum $\hat b_i$ is positive, which makes the matrix strictly diagonally dominant and therefore positive definite. We use the learned $\hat b_i$ for the self capacitance rather than the prior's $C^{\mathrm R}_{ii}$ because the prior's prediction builds up the error of each term in a row; the learned term reduces this error. %

The network comprises two-layer perceptrons of width 64 that encode the conductor vector together with the global vector into $h_i$, and the pair vector together with the mesh step into a pair state $g_{ij}$. Two message-passing rounds form a message on every pair from its two conductors and the pair state, and update every conductor from the mean of the messages of its pairs~\cite{Gilmer2017,Battaglia2018}. After the last round the pair head returns $r_{ij}$ and the node head $q$ reads $h_i$ (Eq.~\eqref{eq:mp}). In physical terms, one round is one sweep of a self-consistent relaxation with the update rule fitted to the data rather than derived, and two rounds let the pair head see, beyond the pair itself, what else each of its two conductors faces. The graph is complete, so the message passing costs $O(N^2)$, a few milliseconds on the layouts of this paper.

\subsection{Models and training}
\label{sec:ml-models}

\begin{figure*}[!tbp]
\centering
yes \includegraphics[width=\textwidth]{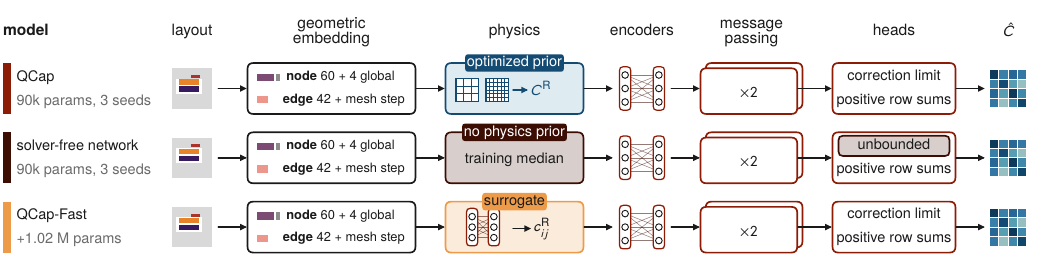}
\caption{The three models of this paper. Each seed of the network has 89858 parameters (1.08\,MB for the three-seed ensemble). QCap-Fast's surrogate adds 1023362 parameters (4.09\,MB), trained on the 346,993 training layouts of the surrogate set, labeled by the prior alone (Appendix~\ref{app:data}).  Note that the physics enters as the base to which the heads add the residual (Eq.~\eqref{eq:prediction}), not as a network input. Only the geometric data is fed to the network.}
\label{fig:model-gallery}
\end{figure*}

The model described so far is the flagship model we release, \emph{QCap}: a three-seed ensemble of networks %
acting on the optimized prior $\CR$, whose outputs we average into a single capacitance matrix. %
This is our most accurate model. We also report two more models: 

The \emph{solver-free} network is the same network as QCap but without the physics prior. Its base is the training median of each pair kind (signal-signal or signal-ground) and its pair head is unbounded, since there is no prior for it to stay close to. This model helps us understand how well the model can perform when harnessing only the richer geometrical representation.

The fast model, \emph{QCap-Fast}, asks whether the physics prior itself can be learned instead of being computed. For this model, we add a second network, trained on an additional 346,993 generated layouts solved by the physics prior only with no Q3D reference (the training part of the 385,593-layout surrogate set of Appendix~\ref{app:data}), to predict the optimized prior $c^{\mathrm R}_{ij}$ from the geometric embedding. QCap's frozen networks are applied unchanged to its output, so this model only affects the physics prior (Appendix~\ref{app:network}). 

We also show the optimized physics prior alone as our baseline reference. All three models share the geometric embedding, the encoders, the message passing, and the closure head, and differ only in the physics they compute at inference: a small correction (QCap), a full solve (solver-free), or a learned stand-in plus a small correction (QCap-Fast) (Fig.~\ref{fig:model-gallery}).

The reference for training is the Q3D matrix $C^{\ast}$ of each training layout whose entries are $C^{\ast}_{ii}=c_{ii}^\ast$ and $C^{\ast}_{ij}=-c_{ij}^\ast$. We use a Huber loss~\cite{Huber1964} on the logarithmic error of every mutual entry, plus down-weighted terms on the self terms (Eq.~\eqref{eq:loss}). Care must be taken in how minibatches are drawn, so that no family is drowned out by the two largest. The parameters are optimized with AdamW~\cite{Loshchilov2019} for a fixed 18,000 steps (Appendix~\ref{app:network}). Each seed has 89,858 parameters and trains in 239\,s on two threads of a GH200 node's CPU. The network accounts for less than 0.2\,s of any prediction, the remainder being the time needed to compute the prior.

\section{Results}
\label{sec:results}

We score the prediction of the model using the following three metrics over its mutual entries $\mathcal S$ (finite positive reference values):

\begin{equation}
\begin{aligned}
\mathrm{med}&=100\operatorname*{median}_{(ij)\in\mathcal S}\frac{|\hat c_{ij}-c^{\ast}_{ij}|}{\max(|c^{\ast}_{ij}|,0.1\fF)},\\
\mathrm{MAE}&=\mathrm{median} \ |\hat c_{ij}-c^{\ast}_{ij}|,\\
\mathrm{RMSE}&=\Big[\frac{1}{|\mathcal S|}\sum_{\mathcal S}(\hat c_{ij}-c^{\ast}_{ij})^2\Big]^{1/2},
\end{aligned}
\label{eq:metrics}
\end{equation}
in \% and fF, averaged over the layouts of the dataset (the \emph{layout mean}), with a 0.1\fF\ floor that keeps attofarad entries at the solver's own noise from dominating a relative error. We call one predictor better than another only when the difference between them is statistically clear, %
otherwise we call them tied (Appendix~\ref{app:protocol}). %
Every time we quote is a wall time from the input of a  layout file to the output of a complete matrix, named with its machine and thread count, i.e., a laptop (Apple M5 Pro), cluster nodes with AMD EPYC 7513 or 7542 CPUs (EPYC nodes) or Intel Xeon E5-2640~v4 CPUs (Xeon nodes), or the Grace CPU of a GH200 node. The test sets of Table~\ref{tab:data} were excluded from training, validation, coefficient fitting and selection, and all networks were frozen before any test set was scored.

\subsection{Performance compared to published models}
\label{sec:res-sota}

\begin{figure*}[!tbp]
\centering
\includegraphics[width=\textwidth]{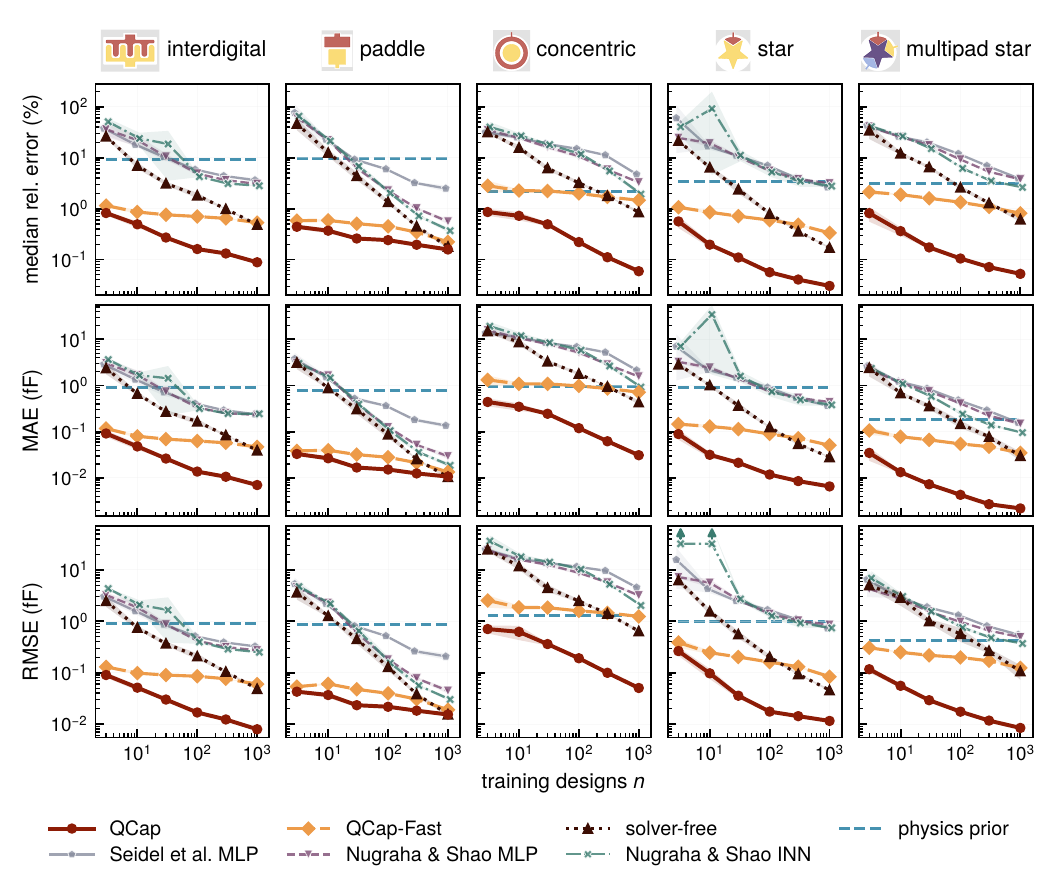}
\caption{Performance of our three models and three literature models when each is trained only on $n$ designs in a single device family, and tested on that family. The physics prior (dashed line) is shown for reference. %
Up arrows indicate INN RMSE off scale, 3469\fF\ ($n=3$) and 3273\fF\ ($n=10$).}
\label{fig:sota}
\end{figure*}

To our knowledge three models predicting capacitance of superconducting quantum devices based on layout have been published: the forward MLP of Seidel et al.~\cite{MLQubitDesign} and the forward MLP and the invertible network (INN) of Nugraha and Shao~\cite{NugrahaShao2023}. We emphasize that these approaches are only able to predict capacitance of the component family they are trained on. We compare performance of models using our QCap and QCap-Fast approaches, but trained only on a single family, against those found in literature. We test on each of the design families in our dataset that have at least 1000 entries, i.e., the five capacitor families of Table~\ref{tab:data-full}. We re-implemented the three published surrogates and trained them on our data, one network per family (Appendix~\ref{app:protocol}). Every predictor was trained from scratch on $n$ designs of a family, with $n$ swept from 3 to 1000 designs. %
The predictions were evaluated on the family's test designs with the results shown in Fig.~\ref{fig:sota} and Table~\ref{tab:sota}.

When only three training designs are used, QCap is 40-170 times more accurate than the published models, and they reach its three-design error only with 700-4100 designs on interdigital and paddle capacitors and never on the other three families. They catch up only on the two families with more than 10,000 training designs, where the Nugraha \& Shao networks trained on the full family are comparable to QCap (Table~\ref{tab:sota}).

Most of the advantage at few designs comes from the inclusion of the physics prior part of our architecture, as demonstrated by lower errors for QCap and QCap-Fast, which include it. QCap-Fast does not improve much with additional training because its floor is mostly the error of its learned prior and not the learned correction. The solver-free network learns faster than QCap in every family but is still 5.5-14.5 times its median error at $n=1000$ on four families, and nearly catches up only on paddle capacitors. QCap is the only model far below the $2\!-\!10\%$ error of the optimized physics prior at every $n$.

\subsection{Generalization at zero shot}
\label{sec:res-zeroshot}

\begin{table*}[!tbp]
\caption{Zero-shot accuracy and inference times of QCap and the physics prior on every test set of Table~\ref{tab:data-full}. Errors are the median relative error in \% (MAE / RMSE in fF). The mean end-to-end wall time is in seconds (colored from green to red), measured on different hardware, and the median Q3D wall time per layout is from our solver records. The marks $^{\ddagger}$, $^{\P}$ and $^{\S}$ indicate details of the timing that are explained in Appendix~\ref{app:protocol}.}
\label{tab:cohorts}
\scriptsize\setlength{\tabcolsep}{1.3pt}
\begin{ruledtabular}
\begin{tabular}{@{}cl@{\hspace{3pt}}cccccc@{}}
\begin{tabular}[b]{@{}c@{}}Sample\\Layout\end{tabular} & Dataset Family (Test Layouts) & \begin{tabular}[b]{@{}c@{}}Q3D\\Median\\(min)\end{tabular} & \begin{tabular}[b]{@{}c@{}}QCap\\\% Error\\(MAE / RMSE, fF)\end{tabular} & \begin{tabular}[b]{@{}c@{}}Physics Prior\\\% Error\\(MAE / RMSE, fF)\end{tabular} & \begin{tabular}[b]{@{}c@{}}EPYC\\2 CPUs (s)\end{tabular} & \begin{tabular}[b]{@{}c@{}}Laptop\\(M5 Pro)\\18 Threads\\(s)\end{tabular} & \begin{tabular}[b]{@{}c@{}}GH200\\32 CPUs\\(s)\end{tabular}\\
\hline
{\raisebox{-0.35\height}{\includegraphics[width=1.3cm,height=0.55cm,keepaspectratio]{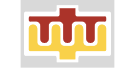}}} & interdigital (1{,}904) & 21.1$^{\ddagger}$ & 0.157 (0.0134 / 0.0164) & 9.332 (0.9270 / 0.8942) & \textcolor[HTML]{2C8A48}{1.13} & \textcolor[HTML]{166631}{0.23} & \textcolor[HTML]{1D7138}{0.38}\\
{\raisebox{-0.35\height}{\includegraphics[width=1.3cm,height=0.55cm,keepaspectratio]{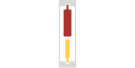}}} & paddle (1{,}533) & 23.9$^{\ddagger}$ & 0.222 (0.0127 / 0.0182) & 9.642 (0.7753 / 0.8709) & \textcolor[HTML]{22793D}{0.539} & \textcolor[HTML]{0B5626}{0.11} & \textcolor[HTML]{12612D}{0.18}\\
{\raisebox{-0.35\height}{\includegraphics[width=1.3cm,height=0.55cm,keepaspectratio]{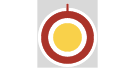}}} & concentric (194) & 34.2$^{\ddagger}$ & 0.171 (0.0862 / 0.1646) & 2.134 (0.957 / 1.306) & \textcolor[HTML]{DD621B}{13.0} & \textcolor[HTML]{389D54}{2.67} & \textcolor[HTML]{3FA85B}{4.34}\\
{\raisebox{-0.35\height}{\includegraphics[width=1.3cm,height=0.55cm,keepaspectratio]{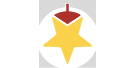}}} & star (188) & 19.6$^{\ddagger}$ & 0.083 (0.0165 / 0.0274) & 3.475 (0.8978 / 0.9776) & \textcolor[HTML]{3FA85B}{4.28} & \textcolor[HTML]{288444}{0.88} & \textcolor[HTML]{2F8F4B}{1.43}\\
{\raisebox{-0.35\height}{\includegraphics[width=1.3cm,height=0.55cm,keepaspectratio]{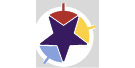}}} & multipad star (190) & 27.5$^{\ddagger}$ & 0.121 (0.0047 / 0.0210) & 3.101 (0.1862 / 0.4297) & \textcolor[HTML]{DE7118}{8.42} & \textcolor[HTML]{30904B}{1.47} & \textcolor[HTML]{359951}{2.21}\\
{\raisebox{-0.35\height}{\includegraphics[width=1.3cm,height=0.55cm,keepaspectratio]{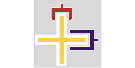}}} & Xmon (578) & 16.7$^{\P}$ & 0.325 (0.0985 / 0.2110) & 8.662 (5.422 / 7.432) & \textcolor[HTML]{2B8947}{1.09} & \textcolor[HTML]{156530}{0.22} & \textcolor[HTML]{1B6F36}{0.34}\\
{\raisebox{-0.35\height}{\includegraphics[width=1.3cm,height=0.55cm,keepaspectratio]{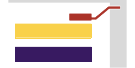}}} & TransmonPocket (171) & 5.2$^{\P}$ & 0.167 (0.0281 / 0.0545) & 2.405 (0.4213 / 0.7196) & \textcolor[HTML]{E08015}{5.39} & \textcolor[HTML]{2A8746}{0.99} & \textcolor[HTML]{30914C}{1.53}\\
{\raisebox{-0.35\height}{\includegraphics[width=1.3cm,height=0.55cm,keepaspectratio]{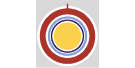}}} & shielded concentric (177) & 11.0$^{\P}$ & 0.157 (0.0615 / 0.1251) & 1.577 (0.631 / 1.431) & \textcolor[HTML]{DB521F}{21.4} & \textcolor[HTML]{3EA75A}{4.18} & \textcolor[HTML]{DF7916}{6.63}\\
{\raisebox{-0.35\height}{\includegraphics[width=1.3cm,height=0.55cm,keepaspectratio]{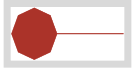}}} & single conductor in a ground cut (147) & 7.0 & 0.594 (2.387 / 2.387) & 13.72 (57.85 / 57.85) & \textcolor[HTML]{30914C}{1.55} & \textcolor[HTML]{1B7037}{0.35} & \textcolor[HTML]{237C3F}{0.60}\\
{\raisebox{-0.35\height}{\includegraphics[width=1.3cm,height=0.55cm,keepaspectratio]{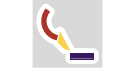}}} & random shape placements (110) & 13.7 & 0.567 (0.0655 / 0.1225) & 11.90 (1.562 / 3.626) & \textcolor[HTML]{21793D}{0.536} & \textcolor[HTML]{0A5425}{0.10} & \textcolor[HTML]{105E2C}{0.16}\\
{\raisebox{-0.35\height}{\includegraphics[width=1.3cm,height=0.55cm,keepaspectratio]{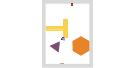}}} & mixed-shape layouts (107) & 15.1 & 0.652 (0.0092 / 0.1584) & 3.957 (0.0753 / 0.7731) & \textcolor[HTML]{E07F15}{5.48} & \textcolor[HTML]{257F41}{0.69} & \textcolor[HTML]{298545}{0.91}\\
{\raisebox{-0.35\height}{\includegraphics[width=1.3cm,height=0.55cm,keepaspectratio]{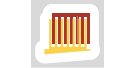}}} & freeform geometries (60) & 20.0 & 0.570 (0.1651 / 0.2873) & 9.112 (2.444 / 4.200) & \textcolor[HTML]{2A8746}{1.01} & \textcolor[HTML]{14632F}{0.20} & \textcolor[HTML]{1A6E36}{0.32}\\
{\raisebox{-0.35\height}{\includegraphics[width=1.3cm,height=0.55cm,keepaspectratio]{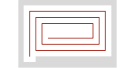}}} & all quantum-metal QComponents (\AtlFinN{}) & 52.0$^{\S}$ & 0.993 (6.615 / 6.631) & 10.77 (84.08 / 84.22) & \textcolor[HTML]{DE7218}{8.18} & \textcolor[HTML]{33954F}{1.87} & \textcolor[HTML]{3BA157}{3.18}\\
{\raisebox{-0.35\height}{\includegraphics[width=1.3cm,height=0.55cm,keepaspectratio]{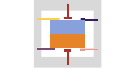}}} & large $N$ (40) & 27.1 & 0.657 (0.0036 / 0.1522) & 1.491 (0.0086 / 0.5106) & \textcolor[HTML]{DC5B1D}{16.1} & \textcolor[HTML]{379B53}{2.45} & \textcolor[HTML]{3CA358}{3.45}\\
{\raisebox{-0.35\height}{\includegraphics[width=1.3cm,height=0.55cm,keepaspectratio]{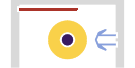}}} & unseen qubits and couplers (116) & 22.5 & 0.672 (0.0936 / 0.3522) & 9.009 (1.875 / 3.654) & \textcolor[HTML]{2E8D4A}{1.32} & \textcolor[HTML]{166631}{0.23} & \textcolor[HTML]{1B7037}{0.35}\\
{\raisebox{-0.35\height}{\includegraphics[width=1.3cm,height=0.55cm,keepaspectratio]{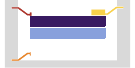}}} & held-out TransmonPockets (36) & 27.2 & 0.326 (0.0189 / 0.1152) & 2.256 (0.0933 / 0.5727) & \textcolor[HTML]{DF7717}{7.00} & \textcolor[HTML]{2D8B49}{1.20} & \textcolor[HTML]{33944E}{1.80}\\
\end{tabular}
\end{ruledtabular}
\end{table*}

The main purpose of QCap is to bypass the simulator. To demonstrate this, we train the model on the full 35,354 simulations of Table~\ref{tab:data-full}. We then run inference using the trained QCap model on all the test datasets in Table~\ref{tab:data-full} with the results shown in Table~\ref{tab:cohorts}. This test set includes device designs and families that the model has never seen in training (i.e. the held-out sets of Table~\ref{tab:data-full}, except for the grids which we discuss in Sec.~\ref{sec:res-windows}).%

On the families it trained on, scored on test designs it never saw, QCap is within 0.08-0.65\% of Q3D (MAE 0.0047-2.39\fF, RMSE 0.016-2.39\fF). On the families it never saw at all, it stays below 1\%, at 0.33-0.99\% (MAE 0.0036-6.62\fF, RMSE 0.115-6.63\fF). It is better than the physics prior on every family and every metric except the MAE of the large-$N$ layouts, where the two are tied: from 2.3 times in median on the large-$N$ layouts (1.49 against 0.657\%; MAE 0.0086 against 0.0036\fF, RMSE 0.511 against 0.152\fF) to 16 times on the freeform geometries (9.11 against 0.570\%; MAE 2.44 against 0.165\fF, RMSE 4.20 against 0.287\fF) among the families it never saw.

Our model is also orders of magnitude faster than the simulator. Q3D takes a median of 5.2-52\,min per layout, while QCap takes 0.54-21\,s on average on two CPUs of an EPYC node (31-2700 times faster), 0.10-4.2\,s on a Macbook laptop (160-13000 times faster) and 0.16-6.6\,s on 32 CPUs of a GH200 node (100-8000 times faster). We timed every test layout on the EPYC nodes and only 12 layouts on all three machines, so the laptop and GH200 times are the EPYC times scaled by the machine ratio fitted on those 12 layouts (Appendix~\ref{app:protocol}). %

\subsection{Scaling beyond the trained domains}
\label{sec:res-scaling}

The training data have two constraints: every simulation has the same substrate (350\um\ thick silicon), and the devices are small. Half of the training layouts are under 0.11\,mm$^2$ (about $0.33\times0.33$\,mm) and 99\% under 5.4\,mm$^2$ with only 244 (0.7\%) of the designs with more than two conductors exceeding $2\times2$\,mm and 15 exceeding $4\times4$\,mm (the largest being a 67-conductor layout on $6.0\times6.6$\,mm). No training layout has more than 68 conductors, and only 137 (0.39\%) have more than 18.

Given these constraints in the training data we cannot expect the model to be highly accurate in these unseen domains. However, since the prior carries the physics of both, we can use that to address the shortcomings of the training datasets.

\subsubsection{Different Substrates}
\label{sec:res-stack}

\begin{figure}[!tbp]
\centering
\includegraphics[width=\columnwidth]{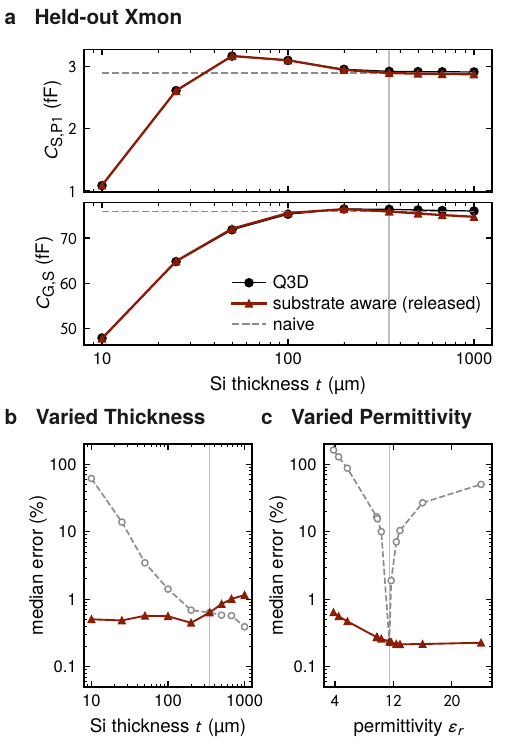}
\caption{Predictions for substrates differing from the training data, which is all on 350\um\ of silicon with $\varepsilon_r=11.45$ (vertical gray lines). The released model computes its physics prior for the substrate the user sets (substrate aware). Without this step (naive), the same networks give their 350\um\ answer on every substrate. (a) For an Xmon that is not in the training data, the coupling between its cross and bus claw, $C_{\mathrm{S,P1}}$, and its cross-to-ground capacitance, $C_{\mathrm{G,S}}$, against silicon thickness. (b) Median error of this Xmon over its mutual entries versus substrate thickness (Table~\ref{tab:stack-doe} gives the means over all 15 designs of this test). (c) Mean median error of the 15 designs at 350\um\ against permittivity.}
\label{fig:substrate}
\end{figure}

To test how well the physics prior corrects for substrate differences, we simulated five designs from each of the three families --- interdigital, shielded concentric and Xmon --- in Q3D at silicon substrate thicknesses from 10 to 1000\um\ ($\varepsilon_r = 11.45$) and at 350\um\ thick substrate permittivities from 3.8 to 24 (Table~\ref{tab:stack-doe}). We compare to a naive approach where the physics prior uses the same 350\um\ silicon substrate for all designs. Figure~\ref{fig:substrate} shows the result for one held-out Xmon against thickness (a, b) and for the 15 designs against permittivity (c). As expected, the naive model returns its 350\um\ answer no matter what the substrate is, so its error grows as the stack moves away from the training one. After including substrate information, the model stays within a few times its training-stack error on every stack tested, while the naive model's error blows up except in cases where the target substrate stack is close to the one seen in training. We also expect the regime at which the naive model fails to depend on the ratio $t/L$ of the substrate thickness to the layout's lateral size, and Fig.~\ref{fig:substrate-tL} shows this dependency, with the naive model's error falling as $(t/L)^{-1.5}$ for $t/L<1$ and the substrate-aware model staying flat throughout (Appendix~\ref{app:res-substrate}). %

\subsubsection{Large Devices}
\label{sec:res-windows}

By construction nothing limits the system size our network can be applied to, since more conductors are just more vectors and the same functions. The method of moments dense solve, however, does not scale, and very few training layouts exist to teach the correction what a large chip looks like. The windowed solve described in Sec.~\ref{sec:emb-prior} ensures that the prior is scalable. %
On a large chip whose ground plane covers everything between the components, the field of the excited conductor is confined by the grounded metal around it. It is a near field across its local gaps and, outside its local neighborhood, approximately the field of a dipole normal to the plane falling as $d^{-3}$. A solve that keeps the neighborhood exactly and simplifies the rest therefore reproduces a global solve to within the mesh error (Appendix~\ref{app:windows}).

\begin{figure*}[!tbp]
\centering
\includegraphics[width=\textwidth]{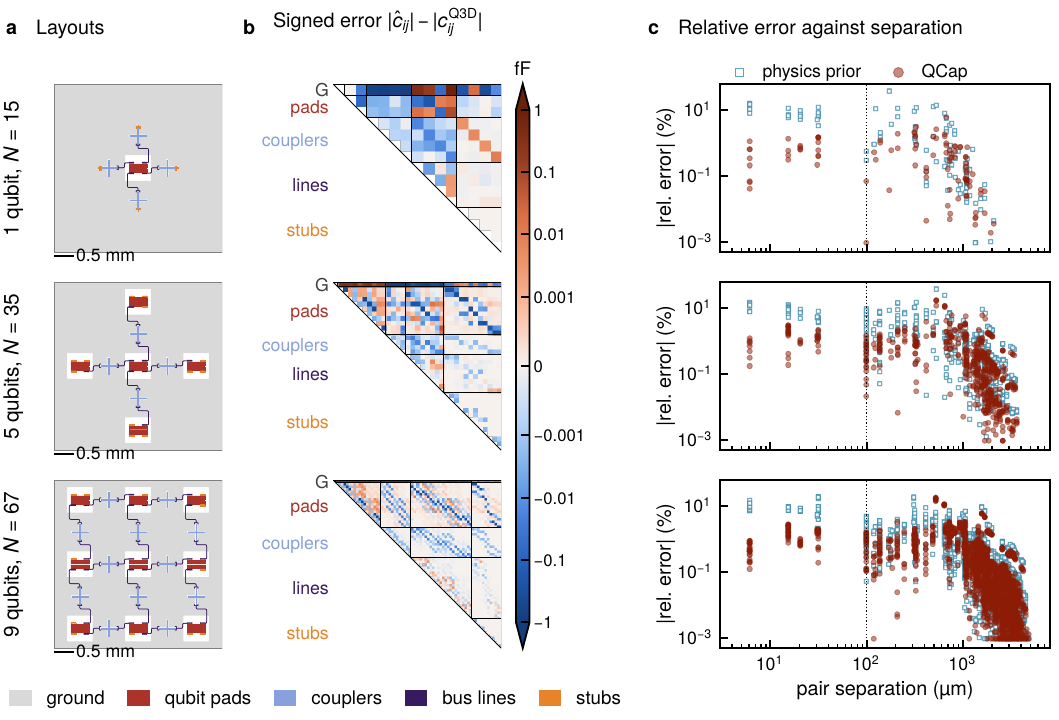}\\
\includegraphics[width=\textwidth]{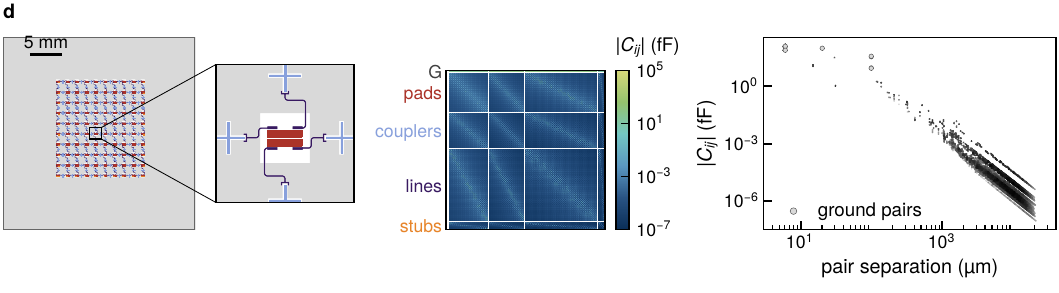}\\
\includegraphics[width=\textwidth]{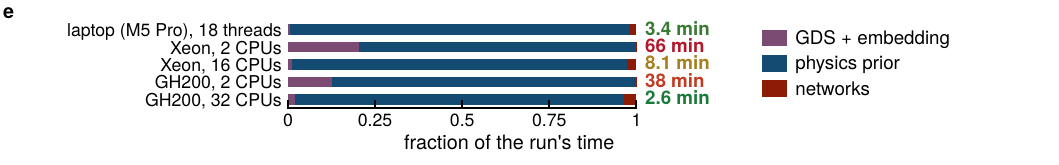}
\caption{QCap on multi-qubit grids, one grid per row in (a)-(c). (a) The one-, five- and nine-qubit grids ($N=15$, 35 and 67), with conductors colored by circuit role. (b) The signed error $|\hat c_{ij}|-|c^{\mathrm{Q3D}}_{ij}|$ of every mutual entry, grouped by role. The seven entries beyond $\pm1$\fF\ are ground capacitances of bus lines and couplers (103-124\fF), each about 1\% low. (c) Relative error of every mutual entry (0.1\fF\ floor) against the pair separation. (d) The 100-qubit chip ($N=781$), for which no field-solver reference exists, with one 1.8\,mm qubit cell enlarged beside it. Its predicted $|C_{ij}|$ is drawn with the ground row and column nine cells wide, next to its 304590 mutual entries against the pair separation, with ground pairs as gray circles. (e) Wall time of QCap on the 100-qubit chip by stage, end to end on different hardware.}
\label{fig:gridscale}
\label{fig:gridfull}
\end{figure*}

We test the model's accuracy at large scale using three devices with multi-qubit grids of two-pad transmons, bus lines, and tunable couplers as shown in Fig.~\ref{fig:gridscale}(a). The devices are one qubit with its four couplers ($N=15$ conductors including ground and floating stubs), a five-qubit cluster ($N=35$), and a $3\times3$ nine-qubit grid ($N=67$), solved in Q3D %
in 12.5, 16.1 and 19.9\,h at 26, 56 and 228\,GB of peak memory (Table~\ref{tab:grid}). The inference times on these three grids from our model was 12.1, 30.3 and 96.5\,s end to end on two CPUs of an EPYC node (3700, 1900 and 740 times faster), 3.4, 6.7 and 11.7\,s on a macbook laptop (13000, 8700 and 6100 times faster) and 2.8, 4.6 and 18.8\,s on 32 CPUs of a GH200 node (16000, 13000 and 3800 times faster), with a peak memory of 2.2-3.3\,GB on two threads of a Xeon node.

On the near pairs QCap is within 0.61, 0.98 and 1.19\% of Q3D on the one-, five- and nine-qubit grids (MAE 0.053, 0.24 and 0.29\fF; RMSE 0.35, 0.43 and 0.42\fF) while on the far pairs beyond 1\,mm its couplings are 0.86-1.13 times Q3D's in median (MAE 0.054-0.10\,aF). For these tiny capacitances we do not put the 0.1 fF floor on the relative error denominator. %
The larger relative error is due both to the very small overall capacitance and to the fact that on these pairs the prior itself sits 1.34-1.47 times above Q3D on the five- and nine-qubit grids (Appendix~\ref{app:fp-far}). This is more than the factor $e^{-\kappa}=0.82$ the network is allowed to remove, so on the far pairs the network applies its maximum correction and the accuracy is limited by the prior. All in all, 2904 of the 2911 mutual entries of the three grids are within 1\fF\ of Q3D. The other seven are the largest ground capacitances of bus lines and couplers (103-124\fF) and are within 1.5\fF\,(about 1\%), which is excellent agreement for systems of this size. %

Lastly, we tested a design %
extending the same grid design to $10\times10$ qubits, giving 100 two-pad transmons at 1.6\,mm pitch and 180 tunable couplers on a $30\times30$\,mm ground plane, $N=781$ conductors. Extrapolating the meshes of the three smaller grids with their length of metal edge, a converged Q3D solution of this chip needs about 48-63 million surface triangles and 2-4\,TB of memory in one shared-memory machine, which is more than the 1.5\,TB of the largest node on our university's clusters, and even on such a machine it would take one week to two months of adaptive passes at 16 cores, with no guarantee of convergence (Appendix~\ref{app:res-grids}). QCap predicts the full $781\times781$ matrix (Fig.~\ref{fig:gridfull}(d)) in 3.4\,min on a laptop (Apple M5 Pro, 18 threads), 2.6\,min on 32 CPUs of a GH200 node and 66\,min on two CPUs of a Xeon node, with a peak resident memory of 2.6-4.4\,GB per process. Most of that time is the windowed physics prior, 79-98\% of the run across these machines %
(Fig.~\ref{fig:gridfull}(e)). Although we make no claims of absolute accuracy here as there is no reference to compare against, this result demonstrates the capability of our approach, allowing us in just a few minutes on a laptop to extract the full capacitance matrix for a 100-qubit chip — a feat beyond the reach of any field solver currently available to us.

\subsection{Practical workflows}
\label{sec:res-workflows}

\begin{figure*}[!t]
\centering
\includegraphics[width=\textwidth]{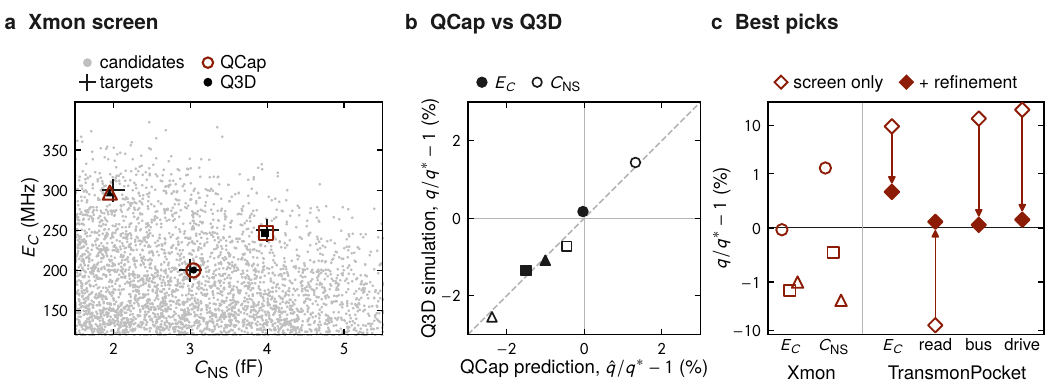}
\caption{(a) 5000 random candidates in the model's ($\CNS$, $E_C$) plane, the three targets (crosses) and, for each target, the candidate the model ranks best as predicted (open red marker) and as solved by Q3D (small black marker). Each target keeps its marker shape in all three panels. (b) Comparison of QCap and Q3D errors versus target for $E_C$ and $\CNS$ of the three Xmons. A point on the dashed diagonal is a design whose model prediction agrees with a Q3D simulation, and its distance from the origin along the diagonal is how far the screened design itself is from the target. $E_C$ is filled and $\CNS$ open. (c) QCap prediction for the best pick of each target, relative to the target. From the random screen alone (open), the picks miss by up to 2.4\% ($-0.048\fF$) for the Xmons and 21\% ($+0.021\fF$) for the TransmonPocket, set by how densely the screen samples each target. Refining the TransmonPocket (filled) brings every quantity within 0.7\%.}
\label{fig:invdes-main}
\end{figure*}

A common workflow is to start with a set of device performance targets and attempt to realize a design to meet those targets. Ideally, we want to use some inverse design~\cite{Molesky2018,Eriksson2026QDO} flows to optimize our layout for our target; however, no publicly known generalized surrogates exist to drive them. Identifying likely parameter relationships and sampling strategies is a major challenge that scales poorly. Since our model can provide accurate results very fast without having to run full simulations, it allows us to bypass this problem by simply running exhaustive screening searches and then confirming the target device behavior with an expensive Q3D solve.

For the first case we attempt to design Xmons at three targets of charging energy and ``claw'' capacitor coupling, $(E_C,\CNS)=(200\MHz,3\fF)$, $(250\MHz,4\fF)$ and $(300\MHz,2\fF)$. We used our design tool to generate 5000 layouts, random within sensible parameter bounds, and ran QCap on all of them. Generating the layouts took about a minute on a laptop and the inference took a few minutes of wall time over 64 parallel processes. The same sweep in Q3D would take hours to days. For each target, we then chose the design candidate that the model predicted would best match and simulated it in Q3D as shown in Fig.~\ref{fig:invdes-main}(a) (the per-target outcomes are in Appendix~\ref{app:m-circuit}). The relatively higher errors for the last two targets are a consequence of the layout sampling not producing a design near the target. Every Xmon point sits on the diagonal of Fig.~\ref{fig:invdes-main}(b), i.e., prediction is within 0.3\% of the Q3D result on every Xmon quantity (at most 0.41\MHz\ discrepancy on $E_C$ and 0.011\fF\ on $\CNS$, the latter on the 4\fF\ target).

Next, we consider a six-conductor TransmonPocket with three couplers and four simultaneous targets, $E_C=200\MHz$ and effective couplings of 10, 1 and 0.1\fF\ to the readout, bus and drive (both defined in Appendix~\ref{app:m-circuit}). This is a problem where it is fairly tough to a priori come up with any parameter space bound. In a similar fashion to before, we again screen 5000 candidates, but this time we follow up with four refinement rounds, doing additional parameter sweeps around the best picks (Appendix~\ref{app:m-circuit}). Solving the final candidate we found in Q3D gives $E_C$ and the readout and bus couplings within 0.5\% of the target ($+0.97\MHz$, $-0.006\fF$ and $-0.0017\fF$ errors) and the drive coupling 1.86\% above it ($+0.0019\fF$ error) (Fig.~\ref{fig:invdes-main}(c)). %
The whole search took about an hour of wall time, 38 min on 32-64 CPUs of our cluster and 21 min on a laptop. In contrast, Q3D simulations of such layouts on 16 cores took a median of 4.8\,min each.

\section{Conclusion}
\label{sec:conclusion}

We have developed a model which predicts the capacitance matrix of a superconducting circuit device. The model uses a geometric embedding that lets one network serve any layout and a cheap physics prior on which the network learns only a bounded residual --- the physics the network would otherwise have to learn from data is computed instead. We achieve percent-level agreement with Ansys Q3D simulations in orders of magnitude less compute time, and are able to predict the capacitance matrix of a (too large to simulate) 100-qubit, 781-conductor chip in 3.4\,min on a laptop.%

The model's accuracy on unseen designs comes from computing the physics at inference rather than during training. This is shown by the alternative solver-free model, which falls well behind its in-family accuracy on out-of-family designs. Likewise, on a family held out of training QCap-Fast fails to capture the physics distilled in its surrogate, while QCap with its computed prior remains accurate. %
Having more (quantity and variation) data especially helps improve the solver-free model (Appendix~\ref{app:res-scaling}) and would improve all models, which is why we strongly request community contributions of such design and simulation datasets to SQuADDS \cite{Shanto2024SQuADDS}. An obvious future direction is to extend to flip-chip and multilayer devices, which will be the subject of future work \cite{rosenberg_solid-state_2020}.

The approach we demonstrate is not specific to capacitance or even electomagnetism. It can be adapted to other domains where design loops exist with an expensive physics solver step. The goal is to create an embedding graph to make sure that the ML model can remain scalable and to add a physics prior that is cheap, contains the physics you want to add, and leaves the network a bounded correction so that the model can also generalize. Future work could develop models for capacitance of non-superconducting circuits, for other electromagnetic quantities, and for other physics simulations such as thermal conductance. 

\begin{acknowledgments}
The authors gratefully acknowledge Firas Abouzahr, Simon Bilodeau, Peter Chen, Enectali Figueroa-Feliciano, Helge Gehring, Madison Howard, Can Knaut, Lionel Martinez, Nicola Pancotti, Taylor Patti, Armin Razavi, Olivia Seidel, Andrew W Senior, Sara Sussman, and Yingheng Tang for useful discussions. The authors gratefully acknowledge NVIDIA for providing the GH200 system used in this work and the National Energy Research Scientific Computing Center (NERSC) for providing compute for benchmarking simulations. Funding was provided by the Army Research Office and the Laboratory for Physical Sciences under grant W911NF-25-1-0255, the National Science Foundation under grant 2612093, and the Avasant Foundation via the SPARQ-AI program.
\end{acknowledgments}

\section*{Statement of AI use}
Generative AI tools (Claude Opus 5.5 and Claude Fable 5.1) were used during the preparation of this manuscript to assist with copyediting and text polishing, and to assist with portions of the analytical derivations and calculations presented in this work. All AI-assisted derivations and calculations were independently verified by the authors; all scientific claims, conclusions, and content of this manuscript remain the sole responsibility of the authors.

\section*{Data and code availability}
All data, code and models in this work are publicly available on Hugging Face~\cite{QCapData,QCapSynthetic,QCapModels} and SQuADDS~\cite{Shanto2024SQuADDS}. Details of access are given in Appendices~\ref{app:ml-usage} and~\ref{app:data}.

\appendix
\section{ML Model Usage}
\label{app:ml-usage}

The models run without installation on a browser as a Hugging Face Space (\url{https://huggingface.co/spaces/SQuADDS/planar-capacitance}). A prediction takes four steps. The user uploads a GDSII file or picks a bundled example. Next, the user verifies that their ground and conductors are identified correctly (we use the convention of layer/datatype 1/0 for ground and 1/10 for all other conductors) and optionally crops the ground or selects a subset of conductors, previewing the prepared layout before solving. The user then chooses the model and the substrate (the silicon stack of the training data by default). Finally, the Space returns the full $N\times N$ Maxwell matrix in fF, ground first and with one row per conductor found in the layout, as a table and as CSV and JSON downloads. For batch screening, users can use the \texttt{gradio\_client}. The Space is a demo that follows standard practice for public tools --- uploads need a sign-in, use and input size are capped per user, and uploaded layouts are not stored. 

For heavy/everyday use, we recommend running the models locally through SQuADDS (version 0.5) as the optional \lstinline|squadds[ml]| extra. The model weights are on Hugging Face (\url{https://huggingface.co/SQuADDS/QCap}). To use the models, users can call the object \lstinline|CapacitancePredictor| and specify the model they want. It takes a GDS path, a \lstinline|gdstk| cell or library, or a list of them, and returns the full Maxwell matrix in fF, ground first (\lstinline|res.names|, \lstinline|res.matrix_fF|), together with the checks for the Maxwell matrix (\lstinline|res.checks|) and the wall time (\lstinline|res.timing_s|). We also ship a shell command \texttt{squadds-ml predict} that does the same without any Python.

For readers who already design with SQuADDS, the model inference is plug-and-play into their present workflows. \lstinline|MLSimulator| has the call and result format of SQuADDS's Q3D simulator, so the ML inference simply replaces the electrostatic solve and \lstinline|Analyzer.predict_capacitance| rebuilds the closest database designs as Quantum Metal components and predicts them. Every prediction converts to the SQuADDS simulation-result format (\lstinline|res.to_sim_results()|), so the Hamiltonian calculators (transmon, Xmon, floating transmon, coupled transmons and resonators) remain compatible. The released datasets can be accessed via the \lstinline|PlanarCapacitanceDB| object. Users can retrain the networks on their own Q3D data and contribute data or models back to SQuADDS. The following code snippet shows how to run inference and do a dataset query using the new API. We have also added detailed tutorials (\url{https://lfl-lab.github.io/SQuADDS/qcap/#tutorials}) to the docsite to go over new workflows that this work enables and how users can train their own models and contribute them back to SQuADDS.

\section{Dataset Availability}
\label{app:data}

\begin{figure*}[p]
\centering
\includegraphics[width=\textwidth]{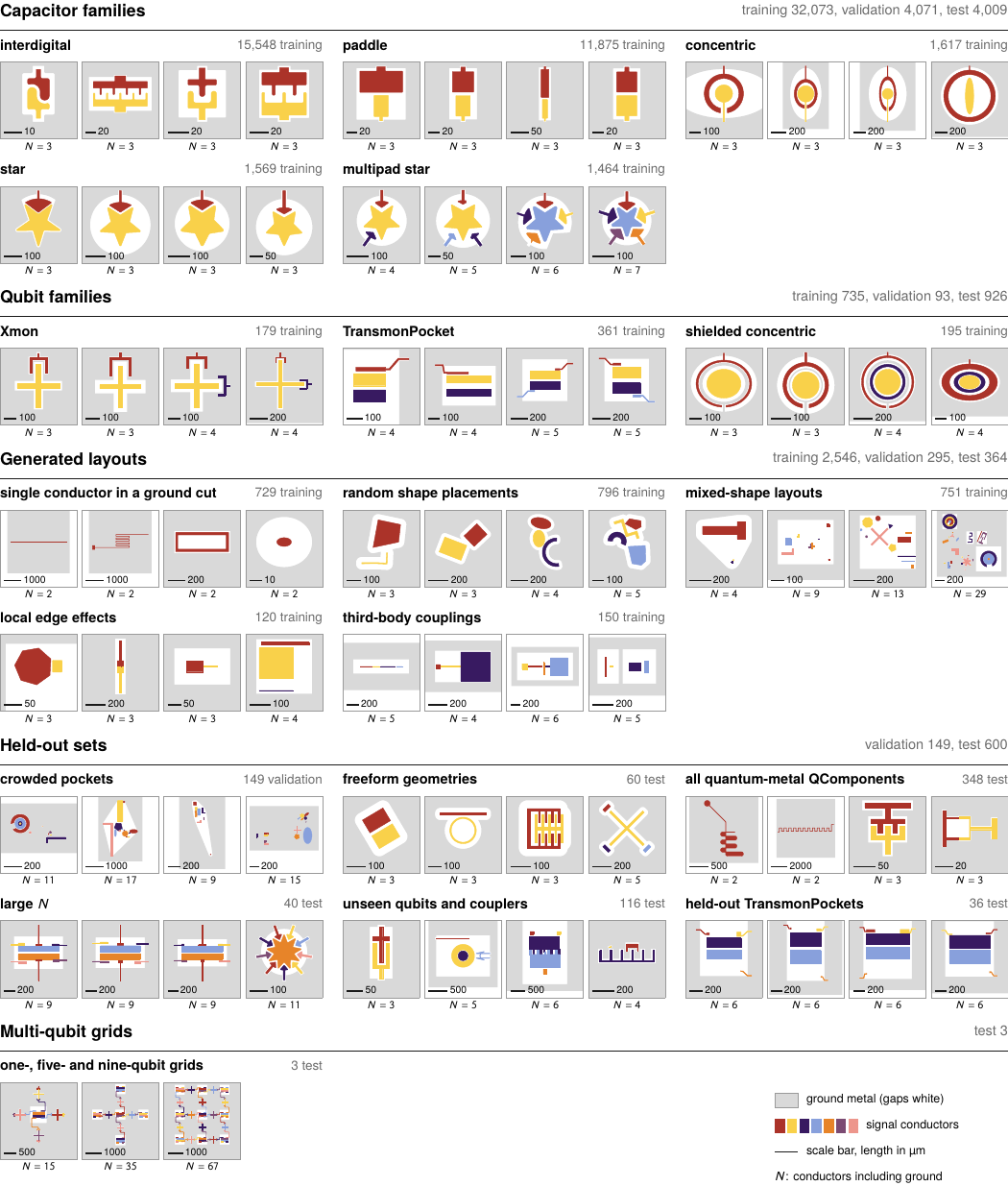}
\caption{Representative layouts of every design family and test set, grouped as in Table~\ref{tab:data} with the splits of each group. Each family shows four layouts, chosen by a fixed sampling rule rather than by hand, with its number of layouts on the right in gray text. Table~\ref{tab:data-full} lists every family.}
\label{fig:dataset-atlas}
\end{figure*}

\begin{table}[!t]
\caption{Every design family %
used in this work, with its splits for training, validation and testing, the number of conductors $N$ including ground, and a representative layout.}
\label{tab:data-full}
\scriptsize\setlength{\tabcolsep}{1.5pt}
\begin{ruledtabular}
\begin{tabular}{@{}>{\raggedright\arraybackslash}p{2.65cm}>{\centering\arraybackslash}p{0.7cm}cccc@{}}
Design Family & $N$ & Training & Validation & Test & Rep. Layout\\
\hline
interdigital & 3 & 15{,}548 & 1{,}931 & 1{,}904 & {\raisebox{-0.35\height}{\includegraphics[width=1.3cm,height=0.55cm,keepaspectratio]{figures/v4/tab_data_thumbs/interdigital.pdf}}}\\
paddle & 3 & 11{,}875 & 1{,}592 & 1{,}533 & {\raisebox{-0.35\height}{\includegraphics[width=1.3cm,height=0.55cm,keepaspectratio]{figures/v4/tab_data_thumbs/paddle.pdf}}}\\
concentric & 3 & 1{,}617 & 170 & 194 & {\raisebox{-0.35\height}{\includegraphics[width=1.3cm,height=0.55cm,keepaspectratio]{figures/v4/tab_data_thumbs/concentric.pdf}}}\\
star & 3 & 1{,}569 & 202 & 188 & {\raisebox{-0.35\height}{\includegraphics[width=1.3cm,height=0.55cm,keepaspectratio]{figures/v4/tab_data_thumbs/star.pdf}}}\\
multipad star & 4--7 & 1{,}464 & 176 & 190 & {\raisebox{-0.35\height}{\includegraphics[width=1.3cm,height=0.55cm,keepaspectratio]{figures/v4/tab_data_thumbs/multipad_star.pdf}}}\\
Xmon & 3--4 & 179 & 18 & 578 & {\raisebox{-0.35\height}{\includegraphics[width=1.3cm,height=0.55cm,keepaspectratio]{figures/v4/tab_data_thumbs/xmon.pdf}}}\\
TransmonPocket & 4--5 & 361 & 48 & 171 & {\raisebox{-0.35\height}{\includegraphics[width=1.3cm,height=0.55cm,keepaspectratio]{figures/v4/tab_data_thumbs/transmonpocket.pdf}}}\\
shielded concentric & 3--5 & 195 & 27 & 177 & {\raisebox{-0.35\height}{\includegraphics[width=1.3cm,height=0.55cm,keepaspectratio]{figures/v4/tab_data_thumbs/shielded_concentric.pdf}}}\\
single conductor in a ground cut & 2 & 729 & 98 & 147 & {\raisebox{-0.35\height}{\includegraphics[width=1.3cm,height=0.55cm,keepaspectratio]{figures/v4/tab_data_thumbs/single_conductor.pdf}}}\\
random shape placements & 3--5 & 796 & 94 & 110 & {\raisebox{-0.35\height}{\includegraphics[width=1.3cm,height=0.55cm,keepaspectratio]{figures/v4/tab_data_thumbs/random_shapes.pdf}}}\\
mixed-shape layouts & 2--68 & 751 & 103 & 107 & {\raisebox{-0.35\height}{\includegraphics[width=1.3cm,height=0.55cm,keepaspectratio]{figures/v4/tab_data_thumbs/mixed_shape.pdf}}}\\
local edge effects & 3--4 & 120 & 0 & 0 & {\raisebox{-0.35\height}{\includegraphics[width=1.3cm,height=0.55cm,keepaspectratio]{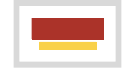}}}\\
third-body couplings & 4--6 & 150 & 0 & 0 & {\raisebox{-0.35\height}{\includegraphics[width=1.3cm,height=0.55cm,keepaspectratio]{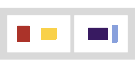}}}\\
\hline
crowded pockets & 6--18 & 0 & 149 & 0 & {\raisebox{-0.35\height}{\includegraphics[width=1.3cm,height=0.55cm,keepaspectratio]{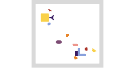}}}\\
freeform geometries & 3--5 & 0 & 0 & 60 & {\raisebox{-0.35\height}{\includegraphics[width=1.3cm,height=0.55cm,keepaspectratio]{figures/v4/tab_data_thumbs/freeform.pdf}}}\\
all quantum-metal QComponents & 2--3 & 0 & 0 & \AtlFinN{} & {\raisebox{-0.35\height}{\includegraphics[width=1.3cm,height=0.55cm,keepaspectratio]{figures/v4/tab_data_thumbs/qcomponents.pdf}}}\\
large $N$ & 8--12 & 0 & 0 & 40 & {\raisebox{-0.35\height}{\includegraphics[width=1.3cm,height=0.55cm,keepaspectratio]{figures/v4/tab_data_thumbs/large_n.pdf}}}\\
unseen qubits and couplers & 3--7 & 0 & 0 & 116 & {\raisebox{-0.35\height}{\includegraphics[width=1.3cm,height=0.55cm,keepaspectratio]{figures/v4/tab_data_thumbs/unseen_qubits.pdf}}}\\
held-out TransmonPockets & 6 & 0 & 0 & 36 & {\raisebox{-0.35\height}{\includegraphics[width=1.3cm,height=0.55cm,keepaspectratio]{figures/v4/tab_data_thumbs/heldout_transmonpocket.pdf}}}\\
one-qubit grid & 15 & 0 & 0 & 1 & {\raisebox{-0.35\height}{\includegraphics[width=1.3cm,height=0.55cm,keepaspectratio]{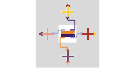}}}\\
five-qubit grid & 35 & 0 & 0 & 1 & {\raisebox{-0.35\height}{\includegraphics[width=1.3cm,height=0.55cm,keepaspectratio]{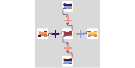}}}\\
nine-qubit grid & 67 & 0 & 0 & 1 & {\raisebox{-0.35\height}{\includegraphics[width=1.3cm,height=0.55cm,keepaspectratio]{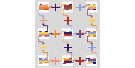}}}\\
\hline
Total & & 35,354 & 4,608 & 5,902 & \\
\end{tabular}
\end{ruledtabular}
\end{table}

All data generated for this paper are available on HuggingFace as two datasets under CC-BY-4.0. The primary dataset, \texttt{SQuADDS\_PlanarCapacitance} (\url{https://huggingface.co/datasets/SQuADDS/SQuADDS_PlanarCapacitance}), contains both the raw data, i.e. every Q3D-solved layout as a GDSII file with its full Maxwell matrix, and the embeddings. The embedding store is structured so that users can query the geometric embedding and the physics prior independently and train their own models. The \texttt{SQuADDS\_PlanarCapacitance-Synthetic} (\url{https://huggingface.co/datasets/SQuADDS/SQuADDS_PlanarCapacitance-Synthetic}) dataset holds generated layouts with no Q3D solution, solved only by the thin-film prior of Appendix~\ref{app:physics}, in two parts, the surrogate set of 385,593 layouts, 346,993 for training and 38,600 for validation, on which QCap-Fast's surrogate physics prior was trained (Appendix~\ref{app:network}) and a generic set of 286,134 layouts, of which 48,075 carry the optimized physics prior $\CR$ of QCap. All of these datasets can be accessed programmatically via the SQuADDS API.

\section{Q3D Simulation Details}
\label{app:q3d}

Every reference matrix in this work is an Ansys Q3D (2025 R2) solution with one set-up: a 350\um\ silicon substrate ($\varepsilon_r=11.45$) on a finite chip in a 1000\um\ vacuum box, zero-thickness conductors and mixed-order basis functions of first and second order, solved for capacitance only at 13\,GHz. Q3D refines its mesh adaptively until every matrix entry changes by less than a target per pass, 0.05\% for three consecutive passes for most families. Before any model was trained, every matrix was also checked against the solver's own integrated surface charge.

\section{Geometric embedding algorithm}
\label{app:embedding}

Section~\ref{sec:embedding} motivated the need for a geometric embedding $\Phi$, stated the  requirements it must satisfy, and introduced our embedding algorithm. Here we provide all the details so that one can recompute them exactly. The full list of the 60 conductor, 42 pair and 4 global columns, with their values for every layout of Table~\ref{tab:data}, is in the three geometric-embedding tables of the published dataset (\url{https://huggingface.co/datasets/SQuADDS/SQuADDS_PlanarCapacitance}), which is also the full list of embedding inputs as the network reads them.

The algorithm reads the metal layers of the GDS file and nothing else, so elements like Josephson junctions, ports, and labels are ignored. Every conductor is a region of the plane and its boundary is a closed contour. The ground is the one exception. Wherever the ground needs a region, for its shape scalars and its width spectrum, we use its etched pocket %
(the complement of the ground within its own bounding box) so the chip outline never enters a conductor vector. All lengths and areas enter the vectors as logarithms in micrometers.

The conductor vector $\mathbf n_i\in\mathbb R^{60}$ has three blocks. The first is 11 shape scalars: a ground flag, log area, log perimeter, compactness, solidity, the radius of the largest inscribed disk, aspect, elongation, the normalized major moment of the area, the number of holes and the mean width. The second block is 33 Fourier magnitudes of the outer contour~\cite{Granlund1972} (Fig.~\ref{fig:walkthrough}(c)). We take the outer contour of the largest part of the region, resample it at $M=512$ points of equal arc length, write the points as complex numbers $z_n=x_n+\mathrm i y_n$ and compute
\begin{equation}
\begin{aligned}
c_k&=\frac1M\sum_{n=0}^{M-1}z_n\,e^{-2\pi\mathrm i kn/M},\\
F_k&=\log\!\big(1+50\,|c_k|/|c_1|\big),\qquad k=-16,\ldots,16,
\end{aligned}
\label{eq:app-fourier}
\end{equation}
with $F_0=0$. The magnitudes are normalized so that they do not depend on position, size, orientation or starting point and describe the shape alone. The fundamental is a circle, the odd harmonics are the corners of a rectangle, and the higher orders are the notches and interdigital teeth. The descriptor is rotation-invariant but not reflection-invariant, because a reflection exchanges $k$ and $-k$, so a mirrored coupler is a different point of the embedding and that invariance has to come from the data. The third block is the width spectrum, the fraction of the area that survives erosion by a disk of radius $r$, at sixteen radii from 0.25 to 100\um. It separates thin strips from compact pads and sets the scale of the fringing field.

The pair vector $\mathbf e_{ij}\in\mathbb R^{42}$ (Fig.~\ref{fig:walkthrough}(c)) is built from the boundary of conductor $i$ that faces conductor $j$ and vice versa. The separation spectrum is the minimum gap and the length of boundary of one conductor that faces the other in each of ten logarithmic distance bins with edges at 0.5, 1.15, 2.63, 6.0, 13.8, 31.6, 72.5, 166, 381, 873 and 2000\um, stored as $\log(1+L_k)$ with $L_k$ the facing length in bin $k$. %
The three coupling moments weight the facing boundary as a coupling would, $\sum d^{-\nu}$ along the facing boundary for $\nu=\tfrac12,1,2$, with the distance $d$ floored at 0.05\um. The conformal coupling applies the capacitance per unit length of two infinite coplanar strips on a half-space~\cite{Simons2001} at every boundary sample. With $d_a$ the distance from sample $a$ of conductor $i$ to the nearest sample of conductor $j$,
\begin{equation}
\chi_{ij}=\frac{\ell}{2}\sum_a\frac{K(k'_a)}{K(k_a)},\qquad
k_a^2=\frac{d_a\,(d_a+w_i+w_j)}{(d_a+w_i)(d_a+w_j)},
\label{eq:app-cps}
\end{equation}
plus the same sum from $j$ to $i$, in units of $\varepsilon_0\epseff$. Here $K$ is the complete elliptic integral of the first kind, $k'=\sqrt{1-k^2}$, $w_i=2A_i/\mathcal P_i$ is the strip width of conductor $i$ with area $A_i$ and perimeter $\mathcal P_i$, and $\ell=0.25\um$ is the boundary sampling step. The distance $d_a$ is floored at 0.05\um\ as for the moments, and for the ground $w\to\infty$ and $k^2=d/(d+w)$. The two shielding shares are the fraction of each conductor's total conformal coupling that this pair carries. The local context is the fractions of ground metal, other metal and etched substrate in rings of 50, 150 and 400\um\ around the conductors. A ground block repeats the separation block against the ground edge within 100\um\ of the conductor only, and a ground flag completes the vector. Every quantity is the mean of the two directions, so the vector is symmetric.

The global vector $\mathbf u\in\mathbb R^4$ holds the four logarithms listed in Sec.~\ref{sec:emb-geometric}: chip area, pocket area, substrate thickness, and effective permittivity. Every entry of the three vectors is in physical units and is standardized by the median and interquartile range of the training data, and these constants are frozen with the model. Laplace's equation has no length scale, so a dilation of the layout shifts every logarithmic length by a constant, which the network can read. The pocket rule for the ground and the 100\um\ ground block are what keep the vectors local.

\section{Optimized Physics Prior Implementation}
\label{app:physics}

Section~\ref{sec:emb-prior} introduced the optimized physics prior $\CR$ and what it does for the model. Here we elaborate on the key concepts for its implementation. Throughout, $p$ is the mesh resolution in \ppg, $p=g/h$ with $g$ the smallest gap of the layout and $h$ the finest panel size, so that the resolution of a solve is one number, and $\mathrm R(p_1,p_2)$ is one Richardson step, Eq.~\eqref{eq:richardson}, from the solves at $p_1$ and $p_2=2p_1$. QCap uses $\mathrm R(1.5,3)$.

\subsection{Integral equation, kernel, panels and solver}
\label{app:substrate}
\label{app:m-prior}

Each entry of the Maxwell matrix is the solution of a boundary-value problem, in which the potential obeys Laplace's equation in the dielectric and takes prescribed values on every conductor, and by Gauss's law the entry is the surface integral of the induced charge, $C_{ij}=\int_{S_i}\sigma^{(j)}dA$. With every conductor a sheet of zero thickness, the surface charge $\sigma^{(j)}$ that puts conductor $j$ at 1\,V and all others at 0\,V obeys the first-kind integral equation~\cite{Harrington1968,Nabors1991FastCap}
\begin{equation}
V_i=\int_{\cup_k S_k}G(\boldsymbol\rho,\boldsymbol\rho')\,\sigma^{(j)}(\boldsymbol\rho')\,dA',\qquad
\boldsymbol\rho\in S_i,\quad V_i=\delta_{ij}.
\label{eq:bem}
\end{equation}
Here $G(\boldsymbol\rho,\boldsymbol\rho')$ is the potential at $\boldsymbol\rho$ of a unit point charge at $\boldsymbol\rho'$ on the substrate surface, and the integral runs over the metal of every conductor. Solving Eq.~\eqref{eq:bem} once per conductor and integrating the charge gives a valid Maxwell capacitance matrix for any planar layout.%

The kernel $G$ is where we encode the substrate information (which is not included in the layout GDS file). For an infinitely thick substrate it is the kernel of a uniform medium of permittivity $\epseff$, $G_\infty(\rho)=1/(4\pi\varepsilon_0\epseff\rho)$. For a substrate of thickness $t$ with vacuum on both sides, which is what the prior assumes everywhere, it is the image series~\cite{Sometani2000}
\begin{equation}
\begin{aligned}
G_t(\rho)&=\frac{1}{4\pi\varepsilon_0\epseff}\left[\frac1\rho+\sum_{n\ge1}\frac{\eta^{2n}+\eta^{2n-1}}{\sqrt{\rho^2+(2nt)^2}}\right],\\
\eta&=\frac{\varepsilon_r-1}{\varepsilon_r+1}.
\end{aligned}
\label{eq:substrate}
\end{equation}
Here $\rho$ is the in-plane distance between the two points and $\eta$ is the reflection coefficient of the dielectric interface. The image at depth $2nt$ is the $n$-th reflection between the two faces of the substrate, and the image weights sum to $\eta/(1-\eta)=\epseff-1$, which makes the far limit exact. Multiplied by $4\pi\varepsilon_0\epseff\rho$, the kernel goes from 1 at $\rho\ll t$ to $\epseff$ at $\rho\gg t$ and reaches the midpoint of the two at $\rho\approx2.7t$, about 1\,mm for 350\um\ of silicon. The transition is this slow because each image pair is weaker than the one before by only $\eta^2$, with $\eta=0.839$ for silicon. We keep the images until $\eta^{2n}<10^{-4}$, 27 pairs for silicon, and add the rest of the weight as one image at that depth. The images are smooth on the scale of a panel and are tabulated once per design, so only the $1/\rho$ term needs exact integrals over the panels, for a panel's own potential and for neighbors closer than 1.5 panel sides.

The substrate kernel is what sets the far couplings of a larger device. On the one-qubit grid of Sec.~\ref{sec:res-windows}, a solve with the half-space kernel $G_\infty$ puts the couplings beyond 1\,mm at 1.80 times the Q3D-simulated values in median, and the substrate kernel at 1.02 times. The kernel covers a single dielectric layer. A multilayer stack would add more interfaces and so more families of images~\cite{Oh1994}, which we have not implemented but is the natural extension of this work.

We discretize Eq.~\eqref{eq:bem} with flat panels. Expanding $\sigma$ in $M$ panels of constant charge $q_m$ and enforcing the equation at the panel centers gives a dense linear system and, from it, the Maxwell matrix of the discretized problem,
\begin{equation}
P\mathbf q=A\mathbf v,\qquad C^{\mathrm{MoM}}=A^{\mathsf T}P^{-1}A .
\label{eq:mom}
\end{equation}
Here $P$ is the panel-to-panel potential matrix, $P_{mn}=|S_n|^{-1}\int_{S_n}G_t(|\boldsymbol\rho_m-\boldsymbol\rho'|)\,dA'$, $A$ is the panel-to-conductor membership matrix with $A_{mi}=1$ if panel $m$ lies on conductor $i$, and $\mathbf v$ is the vector of conductor potentials. Neighboring panels of different sizes make $P$ slightly asymmetric, so we store its near entries as $\tfrac12(P_{mn}+P_{nm})$. $P$ is then symmetric and, in every case we recorded, positive definite, so $C^{\mathrm{MoM}}$ is symmetric and positive definite. %
Up to 12000 panels we factorize $P$ by Cholesky decomposition, and above that a conjugate-gradient solve with a hierarchical-matrix preconditioner takes one right-hand side per conductor. This scheme gives no bound on the energy, so the sign of the mesh error has to be measured rather than derived (Appendix~\ref{app:firstorder}).

The panels are fine at metal edges and narrow gaps and coarser elsewhere, and each region of the mesh is refined only as much as its own geometry needs. They lie on a quadtree attached to the signal conductors, and the finest panel of a cell is $h=\min(g_{\mathrm{loc}}/p,\,g_{\mathrm{ref}}/p)$. Here $g_{\mathrm{loc}}$ is the distance from the cell's metal to the nearest other metal and $g_{\mathrm{ref}}$ is the smallest gap of the layout wider than 0.25\um. A sub-micrometer slit therefore refines only its own neighborhood, and the prior of a component does not change when the chip around it grows.

Panels of size $h$ line every metal edge and are cut along the exact polygon boundary; they double in size away from the edges, and up to 16 collinear panels along a straight edge merge into one strip. The solves at $p=1.5$ and 3 run in a canonical frame, with rotation, origin and conductor order fixed by the geometry, and are averaged over the frames of a symmetric layout, so the mesh does not depend on how the layout was drawn.

\subsection{Mesh error and Richardson step}
\label{app:firstorder}
\label{app:richardson}

Here we give the measurements behind the mesh error of Sec.~\ref{sec:emb-prior} and the reason for the pair $p=1.5$ and 3. On the validation designs the prior lies below the reference on %
98\% of mutual entries at every resolution we solved. We solve $p=0.75$, 1.5 and 3, extracting an error dependence that is roughly linear in $p$---the first five families of Table~\ref{tab:cohorts} have median estimated exponents ranging from 0.84 to 1.18.%

We apply the Richardson step of Eq.~\eqref{eq:richardson} in the logarithm because that keeps the result positive and makes the correction multiplicative, like the network's own. The mesh step of the pair vector is the difference of the same two solves,
\begin{equation}
s_{ij}=\log C^{h}_{ij}-\log C^{2h}_{ij}.
\label{eq:meshstep}
\end{equation}

The choice of the pair is a trade-off between its time and accuracy. On the validation layouts, $\mathrm R(1.5,3)$ roughly halves the error of the single solve at $p=3$, in median, MAE, and RMSE alike, and takes 1.7 times as long. Finer pairs reduce the error further, but slowly and at several times the cost. %
A coarser pair halves the time, but we found the networks trained on it to be less accurate. %
The residual of Eq.~\eqref{eq:residual} is then a small remaining mesh part plus the difference between a thin sheet and the full three-dimensional solve with a finite chip and enclosure.

\subsection{Windows and the far field}
\label{app:windows}
\label{app:fp-far}
\begin{figure*}[!tbp]
\centering
\includegraphics[width=\textwidth]{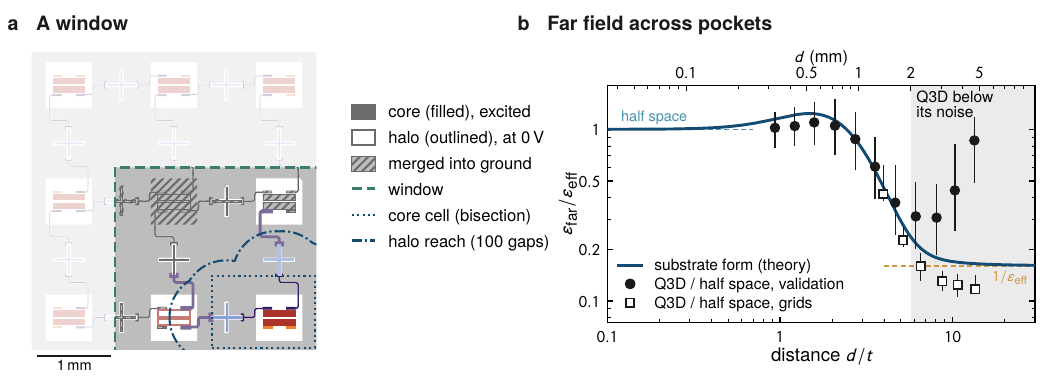}
\caption{The two constructions of the prior for layouts of more than nine conductors. (a) One of the nine windows of the nine-qubit grid. Conductors are colored by circuit role as in Fig.~\ref{fig:gridscale}. Its core conductors (filled), at most eight, are those whose centers fall in one cell of a recursive bisection of the chip (dotted), and they are excited one at a time. The halo, %
every conductor within 100 smallest gaps of the core (dash-dotted), is held at 0\,V. The metal outside the halo but within 100 gaps of the halo conductors is merged into the ground (hatched), and the rest of the metal and gaps are merged into a single filled ground plane. %
The rest of the chip is left out. (b) The substrate factor of Eq.~\eqref{eq:app-far-substrate} for 350\um\ of silicon against the measured ratio of Q3D to the half-space dipole on far pairs across ground pockets, as the median and interquartile range per distance bin. Up to 2\,mm the validation pairs lie at 0.86-1.11 times the curve and the grid pairs at about 0.73 times it. Beyond 2\,mm (shaded) Q3D is below its own noise.}
\label{fig:windows}
\end{figure*}

In large devices we solve the prior in windows. First, the device is bisected into two regions, then each region is bisected, etc.,~until each region forms a \textit{core} of at most eight conductors. Each core gets its own solve window %
(Fig.~\ref{fig:windows}(a)). Each conductor in the core is excited to 1\,V while the others are held at 0\,V. A \textit{halo} bounded by a contour at a distance of 100 smallest gaps from each core conductor defines another region in the window. Conductors which fall inside the halo but not inside the core are always kept at 0\,V. %
All metal that is outside the halo but within 100 gaps of any core or halo conductor is treated as part of the ground plane. Metal which is farther away but inside the window is treated as part of the ground plane, and for this reason all gaps less than 250 $\mu$m are filled with metal. After making these modifications, each window is solved at both mesh levels and Eq.~\eqref{eq:richardson} is applied entry by entry. Two conductors that share no window are always at least 100 gaps apart and take the far field of Eq.~\eqref{eq:far}. The cost of the prior is then a sum over windows plus the far fill, which takes milliseconds.

We checked the locality behind the windows on the 67-conductor grid, against the global solve on the same mesh. The windows reproduce the entries between signal conductors to 0.3-0.6\% (MAE at most 0.014\fF, RMSE at most 0.040\fF) and the ground entries to 1.1\% (0.76 / 1.4\fF). The ground is the least local, because its charge spreads over the whole plane. On the same grid the windows are 4.6 times faster than the global solve on two threads of a Xeon node and 18 times faster on 32 CPUs of a GH200 node, where they run in parallel (Table~\ref{tab:grid-without}). More important than the speed, they bound the memory and time of every solve, which large designs like the 781-conductor chip depend on.

The prior behind every result of this paper, and the one released with the models, was modified from the prior the networks were trained on. There are three modifications, each of which fixes a defect we found on validation layouts of more than nine conductors. First, a conductor too small to hold a single panel used to drop out of its window's solve, leaving its pairs to a rough closed-form estimate; it now gets panels. Second, far pairs inside one ground pocket used a two-dimensional coplanar-strip formula that does not fall off with distance as a real three-dimensional field does; they now take the half-space dipole of Eq.~\eqref{eq:far}. Third, a window's ground plane could be clipped too close to a core conductor that sits far from the ground edge; its margin now widens to reach the ground. None of the three modifications acts on layouts with nine or fewer conductors. We kept the networks trained on the original prior because %
retraining them on the fixed prior made them worse on layouts of a single conductor in a ground cut, whose priors had not changed.

The far field has one fitted constant, the gap share $\alpha$ of the effective area $\tilde A=A+\alpha\mathcal P\bar g$, where $A$ and $\mathcal P$ are the conductor's area and perimeter and $\bar g$ is the mean distance from its boundary to the nearest other metal. We fitted it once on 92 training layouts, before any network was trained, and got $\alpha=0.497$, so the grids test the prior as much as the model.

When the distance between conductors is not small compared to the substrate thickness (as in adjacent ground pockets), the field below the plane sees a substrate of finite thickness, not a half-space. With no backside metal the dipole images alternate in sign, as in the image-dipole series of Ref.~\cite{Martinis2014Coupling}, and Eq.~\eqref{eq:far} holds with $\varepsilon_{\mathrm{far}}(d)=\tfrac12[1+\varepsilon_rR(d)]$ in place of $\epseff$, where
\begin{equation}
R(d)=1+2\sum_{n\ge1}(-\eta)^n\,\frac{1-8n^2t^2/d^2}{(1+4n^2t^2/d^2)^{5/2}},
\label{eq:app-far-substrate}
\end{equation}
with $\eta$ and $t$ as in Eq.~\eqref{eq:substrate}. For $d\ll t$ this is the half-space form, and for $d\gg t$ the coupling drops toward its vacuum value. For 350\um\ of silicon it lowers the coupling between conductors in adjacent ground pockets by about 3$\times$ at the 1.6\,mm qubit pitch of the grids (Fig.~\ref{fig:windows}(b)).

We chose which form applies where by comparing with Q3D. These far couplings are tiny, below one attofarad in median, and beyond 2\,mm are below the Q3D solver noise, so on the validation layouts we compare the pairs between 1 and 2\,mm. Across pockets the half-space form overestimates them about twofold on the validation layouts and sevenfold on the far pairs of the five- and nine-qubit grids, and the substrate form brings them to 0.88 and 1.34-1.47 times Q3D. What remains is at most about an attofarad, MAE 0.022\,aF on the validation pairs and 0.068-0.13\,aF on the grids (RMSE 15.1 and 0.90-0.96\,aF), hundreds of times below QCap's error on the near pairs, and on the grids the network brings these far pairs to 0.86-1.13 times Q3D (Sec.~\ref{sec:res-windows}). Inside one pocket it is the other way round. On the most crowded validation layouts the half-space form matches Q3D (0.97 in median) and the substrate form falls well below it (0.55). The prior therefore uses the half-space form inside a pocket and the substrate form across pockets.

\section{ML Model Architecture and Training Details}
\label{app:network}

Each perceptron of Sec.~\ref{sec:ml-network} is two affine maps with a SiLU nonlinearity between them. The two message-passing rounds $t=1,2$ have separate weights and read
\begin{equation}
\begin{aligned}
m^{(t)}_{ij}&=\mathrm{MLP}^{(t)}_{\mathrm m}\big(h^{(t-1)}_i+h^{(t-1)}_j,\\
&\hphantom{=\mathrm{MLP}^{(t)}_{\mathrm m}\big(}|h^{(t-1)}_i-h^{(t-1)}_j|,\ g_{ij}\big),\\
h^{(t)}_i&=h^{(t-1)}_i+\mathrm{MLP}^{(t)}_{\mathrm u}\Big(h^{(t-1)}_i,\ \tfrac{1}{N-1}\sum\nolimits_{j\neq i}m^{(t)}_{ij}\Big),
\end{aligned}
\label{eq:mp}
\end{equation}
so the message is symmetric in $i\leftrightarrow j$~\cite{Gilmer2017,Battaglia2018}. After the last round the pair head reads $(h_i+h_j,|h_i-h_j|,g_{ij})$, which makes $r_{ij}=r_{ji}$. Besides the properties of Sec.~\ref{sec:learning}, the form of Eq.~\eqref{eq:prediction} makes the signal block of $\hat C$, the matrix a circuit is extracted from, positive definite as well. Relabeling the conductors permutes $\hat C$ in the same way, and the number of parameters does not depend on $N$.

\label{app:kappa}The knob $\kappa$ must cover the prior's typical error, or the network cannot correct it, but not much more, or the bound stops protecting the prediction where the data run out. Our $\kappa=0.2$ is therefore an empirical value, set once for our chosen optimized physics prior; a coarser or finer prior would need a different one. We also let the network learn $\kappa$, as one value for all entries, one per pair kind or one per entry. Learned freely, it rises to about 0.4, but we see no performance benefit to the overall network. Since this learning takes %
extra training time, we just keep $\kappa$ at the empirical value as explained in Sec.~\ref{sec:ml-models}.

The loss of Sec.~\ref{sec:ml-models} over a minibatch $\mathcal B$ is
\begin{equation}
\begin{aligned}
\mathcal L&=\frac{1}{|\mathcal M_{\mathcal B}|}\sum_{(ij)\in\mathcal M_{\mathcal B}}
\mathrm H_{\delta}\!\big(\log\hat c_{ij}-\log c^{\ast}_{ij}\big)\\
&\quad+0.1\,\overline{\mathrm H_{\delta}}\big(\log\hat C_{ii}-\log C^{\ast}_{ii}\big)_{i\in\mathrm{signal}}\\
&\quad+0.02\,\mathrm H_{\delta}\big(\log\hat C_{00}-\log C^{\ast}_{00}\big),\\
\mathrm H_{\delta}(x)&=\begin{cases}\tfrac12x^2&|x|\le\delta\\ \delta\,(|x|-\tfrac12\delta)&|x|>\delta,\end{cases}
\qquad\delta=0.05,
\end{aligned}
\label{eq:loss}
\end{equation}
where $\mathcal M_{\mathcal B}$ is the set of mutual entries with a finite positive reference and the bar denotes the mean. The logarithm makes the loss relative, so that it is insensitive to the absolute value of the capacitance, e.g., a 70\,aF coupling and a 100\fF\ ground capacitance count alike. The Huber threshold keeps the few entries with errors above 5\%, mostly small couplings at the chip margin, from setting the gradient. The ground self term is weighted down rather than left out, because it depends on the simulation box of the reference and the closure still needs a finite target.

Each step draws one family uniformly, then a conductor-count shard of that family in proportion to its layout count, then $\min(64,\lfloor8192/\text{pairs}\rfloor)$ layouts, so a 68-conductor layout is never padded against three-conductor capacitors. AdamW runs at a learning rate of $10^{-3}$ with weight decay $10^{-4}$, a cosine schedule to $10^{-5}$ and gradient norms clipped at 5, with no early stopping, and every standardized input (Appendix~\ref{app:embedding}) is clipped to $\pm10$. Every fit in the paper uses these settings and takes a few minutes, depending on the machine and on how many fits share it. The solver-free network has the same size and settings. The three seeds of QCap were selected on the validation set and frozen before any test set was read.

QCap-Fast's surrogate is a larger network of the same kind, with four message-passing rounds of width 128 and eleven times the parameters. It learns $c^{\mathrm R}_{ij}$ from the geometric embedding on the surrogate set of Appendix~\ref{app:data}, generated layouts that cost no field-solver time and cover the training geometry of every family. It trains for 60000 steps, about half an hour on the GPU of a GH200 node, and inference runs in 5.6\,ms per layout.

\section{Evaluation Details}
\label{app:protocol}

We call two predictors tied unless a paired layout bootstrap (10000 resamples) over the same layouts and entries gives a 95\% interval that excludes zero and a point estimate larger than 0.25 percentage points of median error or 0.05\fF\ of MAE. %

The EPYC times of Table~\ref{tab:cohorts} were measured on EPYC nodes except for 3 of 61 timing batches, which ran on a Xeon Silver 4116 node. The laptop and GH200 times are estimated from them with a machine ratio fitted on 12 layouts timed on all three machines (26\% RMS residual).$^{\ddagger}$ For the first five families of Table~\ref{tab:data-full}, interdigital to multipad star, the Q3D time is the median wall time of their training-data solves. These ran on 20 cores of our cluster, mostly on EPYC nodes (64-87\% of each family's designs) and otherwise on Xeon Silver 4116 nodes or across nodes of different CPUs, except 407 interdigital designs solved on 14 cores of a lab PC (AMD Threadripper PRO 3955WX) and 113 designs solved on one core. $^{\P}$No solve times were recorded for Xmon, TransmonPocket and shielded concentric, and their Q3D time is the median of our own re-solves of ten designs per family on five substrate stacks, on one core. $^{\S}$The Q3D time is over the 281 of the \AtlFinN{} QComponents that were solved on eight cores of our cluster.

Of the three published surrogates of Sec.~\ref{sec:res-sota}, the forward MLP of Seidel et al.~\cite{MLQubitDesign} follows the authors' released code. The forward MLP and the INN, read in the forward direction, of Nugraha and Shao~\cite{NugrahaShao2023} use assumed hyperparameters, since neither their settings nor their code is public, and for these two we tested two configurations before any fit and report the one with the lower validation error over all families. Each network is adapted to predict the mutual entries, with a masked loss over the conductor slots a design has and early stopping on the family's validation split. At every $n$ of Fig.~\ref{fig:sota}, all predictors share the same three draws of designs and seeds.

We also checked the in-family test split for near duplicates of training layouts, and no test layout duplicates a training geometry. The test designs are a random split of gridded parameter sweeps, so a typical test design is one grid step away from its nearest training design in one or a few parameters, as the validation designs are. Leaving out the quarter of the test designs closest to training changes the in-family error by less than 0.01 percentage points, and its MAE and RMSE by at most 0.002\fF. Only on concentric, the family with the fewest training designs, does the error grow with the distance to training, by about 1.6 times in median in the farthest fifth.

\begin{table*}[!tbp]
\caption{In-family comparison of the published parameter-based surrogates, the physics prior and QCap, per family at the full training set. Errors are the median relative error in \% (MAE / RMSE in fF). Median time per design is for the parameter networks on one thread and, end to end, for the physics prior and QCap on two CPUs of an EPYC node.}
\label{tab:sota}
\scriptsize\setlength{\tabcolsep}{3pt}
\begin{ruledtabular}
\begin{tabular}{lccccc}
Family (Training Designs) & Seidel et al.\ MLP & Nugraha \& Shao MLP & Nugraha \& Shao INN & Physics Prior & QCap\\
\hline
interdigital (15,548) & 2.898 (0.215 / 0.308) & 0.147 (0.011 / 0.013) & 0.091 (0.0061 / 0.008) & 9.332 (0.927 / 0.894) & 0.157 (0.013 / 0.016)\\
paddle (11,875) & 1.215 (0.068 / 0.098) & 0.183 (0.010 / 0.015) & 0.154 (0.0084 / 0.014) & 9.642 (0.775 / 0.871) & 0.222 (0.013 / 0.018)\\
concentric (1,617) & 3.025 (1.326 / 3.305) & 1.805 (0.843 / 1.894) & 1.130 (0.562 / 1.335) & 2.134 (0.957 / 1.306) & 0.171 (0.086 / 0.165)\\
star (1,569) & 1.831 (0.253 / 0.514) & 2.431 (0.334 / 0.652) & 2.255 (0.331 / 0.626) & 3.475 (0.898 / 0.978) & 0.083 (0.017 / 0.027)\\
multipad star (1,464) & 2.920 (0.109 / 0.420) & 2.846 (0.099 / 0.380) & 1.995 (0.072 / 0.294) & 3.101 (0.186 / 0.430) & 0.121 (0.0047 / 0.021)\\
\hline
time per design & 0.21\,ms & 0.02\,ms & 0.27\,ms & 0.075\,s  & 0.755\,s \\
\end{tabular}
\end{ruledtabular}
\end{table*}

\section{Supporting Results}
\label{app:results}

\subsection{Family hold-out}
\label{app:sf-lofo}

To test the model on a family it has never seen but that resembles its training data, we remove one family at a time from the training data, train the same network on the rest and score it on the validation designs of the removed family. We do this for the five families with the most training designs, i.e., interdigital, paddle, concentric, star and multipad star, with three seeds each. The family is removed from everything the model learns from, including the standardization of its inputs. Table~\ref{tab:sf-shipped} gives the results. Without the family, the model's error is about 2.4 times its error when trained on all families, and still about 14 times lower than that of the prior. Removing a family by name does not remove every similar geometry, though. The shielded concentric designs, for example, stay in training when concentric is removed.

QCap-Fast behaves differently, because its learned prior is retrained without the family too. On families that the learned prior's training data cover it is at 0.606\% (MAE 0.079\fF, RMSE 0.128\fF), but on a removed family it is at 3.113\% (0.449 / 0.633\fF), worse than the physics it learns to imitate. Adding the 48075 generated layouts of the generic set (Appendix~\ref{app:data}) to the learned prior's training improves this by only about a tenth. A learned prior needs layouts of the new family itself, and those are the data our community is short of, which is why contributions to shared design databases such as SQuADDS matter (Sec.~\ref{sec:conclusion}).

\begin{table*}[!tbp]
\caption{QCap's network trained without one of the five largest families of Table~\ref{tab:data} and scored on the validation designs of that family, averaged over three seeds. Errors are the median relative error in \% (MAE / RMSE in fF). The physics prior is shown for comparison. Only the last column was trained on the family.}
\label{tab:sf-shipped}
\scriptsize\setlength{\tabcolsep}{1.5pt}
\begin{ruledtabular}
\begin{tabular}{lccc}
Family (Layouts) & Physics Prior & Trained without the Family & Trained on All Families\\
\hline
interdigital (1931) & 9.282 (0.931 / 0.895) & 0.413 (0.0392 / 0.0435) & 0.151 (0.0132 / 0.0163)\\
paddle (1592) & 9.668 (0.770 / 0.868) & 0.417 (0.0269 / 0.0312) & 0.210 (0.0120 / 0.0180)\\
concentric (170) & 2.312 (1.085 / 1.361) & 0.544 (0.2423 / 0.4906) & 0.176 (0.0816 / 0.1334)\\
star (202) & 3.360 (0.868 / 0.926) & 0.200 (0.0308 / 0.0492) & 0.083 (0.0160 / 0.0252)\\
multipad star (176) & 3.037 (0.188 / 0.446) & 0.376 (0.0125 / 0.0771) & 0.119 (0.0056 / 0.0216)\\
\hline
average of the five families & 5.532 (0.769 / 0.899) & 0.390 (0.0703 / 0.1383) & 0.148 (0.0257 / 0.0429)\\
all 4071 designs together & 8.578 (0.840 / 0.886) & 0.408 (0.0413 / 0.0591) & 0.171 (0.0154 / 0.0225)\\
\end{tabular}
\end{ruledtabular}
\end{table*}

\subsection{Substrates}
\label{app:res-substrate}

\begin{figure}[!tbp]
\centering
\includegraphics[width=\columnwidth]{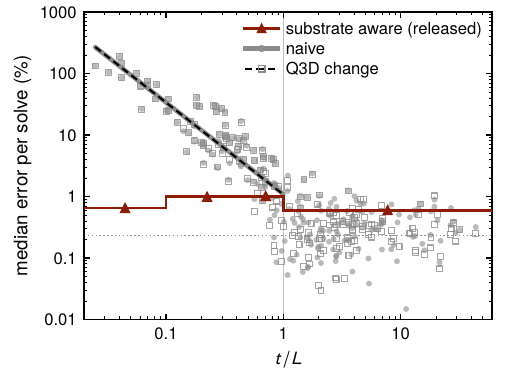}
\caption{Error on the silicon solves of the substrate study (Sec.~\ref{sec:res-stack}) against $t/L$, the substrate thickness over the lateral size of the layout. Each point is the median error of one solve. For $t/L<1$ the naive error (filled circles) follows how far the Q3D matrix itself moves from its 350\um\ solution (open squares, power-law fits), while the substrate-aware released model, shown as layout means in four bins of $t/L$, stays flat. The dotted line is the error on the training substrate.}
\label{fig:substrate-tL}
\end{figure}

Table~\ref{tab:stack-doe} shows the results of the substrate study of Sec.~\ref{sec:res-stack}. On silicon the error of the naive model, which assumes the substrate is unchanged, closely tracks the change of the Q3D matrix itself from its 350\um\ solution. %
Figure~\ref{fig:substrate-tL} plots these silicon solves against $t/L$, the substrate thickness over the lateral size of the layout. For $t/L\ge1$ the released model keeps its training-stack accuracy.

\begin{table*}[!tbp]
\caption{QCap on substrates other than its training stack, over the 15 designs of Sec.~\ref{sec:res-stack}. Errors are the median relative error in \% (MAE / RMSE in fF). Naive and substrate aware are as in Fig.~\ref{fig:substrate}, and the last column is how far Q3D itself moves from its 350\um\ silicon solution. For the other materials that solution is first rescaled by the change of $\epseff$, so the column is the change beyond a plain permittivity rescaling.}
\label{tab:stack-doe}
\scriptsize\setlength{\tabcolsep}{3pt}
\begin{ruledtabular}
\begin{tabular}{>{\raggedright\arraybackslash}p{0.18\textwidth}>{\centering\arraybackslash}p{0.2\textwidth}>{\centering\arraybackslash}p{0.2\textwidth}>{\centering\arraybackslash}p{0.2\textwidth}}
Substrate & Naive & Substrate Aware (Released) & Q3D Change from 350\um\ Si\\
\hline
Si 10\um & 76.64 (8.566 / 18.789) & 0.69 (0.077 / 0.164) & 76.46 (8.545 / 18.742)\\
Si 25\um & 14.58 (4.157 / 9.132) & 0.93 (0.174 / 0.290) & 14.59 (4.149 / 9.093)\\
Si 50\um & 5.88 (2.038 / 4.298) & 1.02 (0.187 / 0.353) & 5.72 (2.042 / 4.265)\\
Si 100\um & 2.88 (0.960 / 1.660) & 0.92 (0.189 / 0.351) & 3.02 (0.970 / 1.636)\\
Si 200\um & 0.70 (0.232 / 0.359) & 0.58 (0.141 / 0.261) & 0.69 (0.222 / 0.311)\\
Si 350\um\ (training) & 0.23 (0.053 / 0.111) & $=$ naive & 0\\
Si 1000\um & 0.50 (0.134 / 0.245) & 0.65 (0.217 / 0.451) & 0.36 (0.080 / 0.183)\\
fused silica, $\varepsilon_r=3.8$ & 162.9 (20.47 / 30.50) & 0.64 (0.058 / 0.093) & 1.31 (0.136 / 0.290)\\
sapphire, 10.34 & 10.0 (3.00 / 4.48) & 0.26 (0.050 / 0.106) & 0.10 (0.022 / 0.047)\\
Si, room temperature, 11.7 & 1.89 (0.630 / 0.942) & 0.23 (0.054 / 0.113) & 0.02 (0.0049 / 0.011)\\
GaAs, 12.9 & 10.4 (3.84 / 5.72) & 0.21 (0.057 / 0.120) & 0.11 (0.030 / 0.064)\\
Ge, 16.0 & 26.9 (12.17 / 18.09) & 0.22 (0.081 / 0.158) & 0.30 (0.120 / 0.218)\\
LaAlO$_3$, 24 & 50.4 (33.61 / 50.01) & 0.23 (0.115 / 0.219) & 0.56 (0.318 / 0.603)\\
\end{tabular}
\end{ruledtabular}
\end{table*}

\subsection{Large Qubit Grids}
\label{app:res-grids}

\begin{table*}[!tbp]
\caption{Errors and inference times of our models against Q3D on the multi-qubit grids. Errors are the median relative error in \% (MAE / RMSE in fF). The last three columns are wall times on the hardware named in their headings, end to end for QCap and the solver-free network, while the physics-prior time is the time of its two solves within QCap's run. The Q3D times are on the number of cores of one cluster node given in their row. $^{\ddagger}$Timed only in the paired runs of Table~\ref{tab:grid-without}. $^{\dagger}$Converged to 0.062\% instead of 0.05\% because of out-of-memory errors.}
\label{tab:grid}
\scriptsize\setlength{\tabcolsep}{1pt}
\begin{ruledtabular}
\begin{tabular}{llccccccc}
$N$ & Predictor & All Entries & $\ge0.1$\fF & Within 100\um & Beyond 100\um & \begin{tabular}[b]{@{}c@{}}EPYC\\2 CPUs\end{tabular} & \begin{tabular}[b]{@{}c@{}}Laptop\\(M5 Pro)\\18 Threads\end{tabular} & \begin{tabular}[b]{@{}c@{}}GH200\\32 CPUs\end{tabular}\\
\hline
15 & Q3D & 105 entries & 45 entries & 27 entries & 78 entries & \multicolumn{3}{c}{\textcolor[HTML]{840924}{12.5\,h} on 1 core}\\
 & QCap & 0.370 (0.00074 / 0.1769) & 0.648 (0.0156 / 0.2702) & 0.612 (0.0532 / 0.3487) & (0.00027 / 0.0034) & \textcolor[HTML]{2A8746}{12.1\,s} & \textcolor[HTML]{186B34}{3.37\,s} & \textcolor[HTML]{166631}{2.78\,s}\\
 & global solve$^{\ddagger}$ & 0.520 (0.0010 / 0.2021) & 0.760 (0.0127 / 0.3087) & 0.761 (0.2558 / 0.3985) & (0.00046 / 0.0029) & -- & -- & --\\
 & physics prior & 0.871 (0.0016 / 3.168) & 7.094 (0.088 / 4.839) & 7.805 (1.541 / 6.247) & (0.00024 / 0.0301) & \textcolor[HTML]{2A8645}{11.5\,s} & -- & --\\
 & solver-free & 2.271 (0.0033 / 3.212) & 12.07 (1.369 / 4.906) & 11.42 (1.953 / 6.331) & (0.0016 / 0.0988) & \textcolor[HTML]{02461D}{0.67\,s} & -- & --\\
\hline
35 & Q3D & 595 entries & 123 entries & 67 entries & 528 entries & \multicolumn{3}{c}{\textcolor[HTML]{7F0823}{16.1\,h} on 16 cores}\\
 & QCap & 0.209 (0.00025 / 0.1448) & 0.952 (0.0159 / 0.3185) & 0.975 (0.2420 / 0.4315) & (0.00016 / 0.0031) & \textcolor[HTML]{379C53}{30.3\,s} & \textcolor[HTML]{227A3D}{6.65\,s} & \textcolor[HTML]{1D7238}{4.61\,s}\\
 & global solve$^{\ddagger}$ & 0.249 (0.00027 / 0.1825) & 1.208 (0.0238 / 0.4013) & 1.609 (0.2969 / 0.5437) & (0.00016 / 0.0031) & -- & -- & --\\
 & physics prior & 0.340 (0.00035 / 1.627) & 4.992 (0.085 / 3.577) & 6.668 (1.162 / 4.847) & (0.00027 / 0.0104) & \textcolor[HTML]{369A52}{28.1\,s} & -- & --\\
 & solver-free & 1.948 (0.0022 / 2.736) & 15.74 (0.409 / 6.017) & 13.22 (1.457 / 8.151) & (0.0015 / 0.0573) & \textcolor[HTML]{12612D}{2.17\,s} & -- & --\\
\hline
67 & Q3D & 2{,}211 entries & 257 entries & 135 entries & 2{,}076 entries & \multicolumn{3}{c}{\textcolor[HTML]{7B0622}{19.9\,h}$^{\dagger}$ on 32 cores}\\
 & QCap & 0.105 (0.00011 / 0.1041) & 0.872 (0.0183 / 0.3053) & 1.186 (0.2940 / 0.4211) & (0.000089 / 0.0022) & \textcolor[HTML]{DF7816}{96.5\,s} & \textcolor[HTML]{2A8646}{11.7\,s} & \textcolor[HTML]{31914C}{18.8\,s}\\
 & global solve$^{\ddagger}$ & 0.113 (0.00012 / 0.1191) & 1.119 (0.0219 / 0.3494) & 1.294 (0.2802 / 0.4820) & (0.000097 / 0.0025) & -- & -- & --\\
 & physics prior & 0.170 (0.00017 / 1.267) & 4.125 (0.070 / 3.716) & 7.918 (1.077 / 5.128) & (0.00015 / 0.0036) & \textcolor[HTML]{DF7A16}{91.1\,s} & -- & --\\
 & solver-free & 1.304 (0.0013 / 2.848) & 18.40 (0.432 / 8.353) & 16.60 (1.990 / 11.52) & (0.0012 / 0.0471) & \textcolor[HTML]{1F753B}{5.47\,s} & -- & --\\
\end{tabular}
\end{ruledtabular}
\end{table*}

As seen from Table~\ref{tab:grid-without}, filling the far pairs with the half-space dipole instead of the substrate form leaves every entry within 100\um\ and every entry above 0.1\fF\ unchanged, but multiplies the MAE of the far pairs by about 8 and 12 on the five- and nine-qubit grids. One global solve instead of the windows makes the same networks less accurate in almost every scope of every grid and 3 to 18 times slower on the GH200 node. It is less accurate because the networks were trained on windowed priors, whose entries within 100\um\ differ from the global ones by 0.17-0.47\% (MAE 0.043-0.079\fF, RMSE 0.102-0.215\fF), and slower because the windows run in parallel and one global solve does not. Without any prior, the solver-free network is more than five times less accurate than QCap on the near pairs. 

The estimate for the Q3D solution time of simulating the 100-qubit chip in Sec.~\ref{sec:res-windows} is made by scaling the converged meshes of the three grids with the length of metal edge, 2221\,mm on the chip, whose smallest gap is 6\um. In the prediction, 100 windows per mesh level carry the solves, 294628 of the pairs take the far field of Eq.~\eqref{eq:far} in about two seconds per level, and from 16 CPUs up the embedding runs during the solves. The predicted matrix is symmetric to machine precision with every mutual entry negative, and its entries run from 124\fF\ for adjacent metal to $9\times10^{-8}\fF$ across the chip. It is the same on the cluster nodes to $5.5\times10^{-7}$ relative, and the laptop, whose meshes are not bit-compatible, agrees with them to 0.011\fF\ at most, with most entries bit-identical. The three seeds of the network disagree by 0.98\% in median on the 1640 pairs within 100\um\ (MAE 0.090\fF, RMSE 0.81\fF) and not at all beyond 1\,mm, where all three apply the full correction of $\pm20\%$ that $\kappa$ allows.

\begin{table*}[!tbp]
\caption{The grids of Table~\ref{tab:grid} with the physics prior changed. Global solve replaces the windows with one solve of the whole layout, half-space fills the far pairs without the substrate factor, the solver-free network has no prior and QCap-Fast has a learned one. Errors are the median relative error in \% (MAE / RMSE in fF). Times are end to end, with the prior's share in parentheses, from paired runs separate from those of Tables~\ref{tab:cohorts} and \ref{tab:grid}. Peak is the resident memory on the Xeon node. $^{\ast}$Embedding only, on two CPUs of an EPYC node.}
\label{tab:grid-without}
\tiny\setlength{\tabcolsep}{1pt}
\begin{ruledtabular}
\begin{tabular}{llccccccc}
$N$ & Predictor & All Entries & $\ge0.1$\fF & Within 100\um & Beyond 100\um & \begin{tabular}[b]{@{}c@{}}Xeon\\2 Threads\end{tabular} & \begin{tabular}[b]{@{}c@{}}GH200\\32 CPUs\end{tabular} & \begin{tabular}[b]{@{}c@{}}Peak\\(GB)\end{tabular}\\
\hline
15 & QCap, windows & 0.370 (0.00074 / 0.1769) & 0.648 (0.0156 / 0.2702) & 0.612 (0.0532 / 0.3487) & (0.00027 / 0.0034) & 17.2\,s (15.8) & 3.1\,s (2.4) & 2.19\\
 & QCap, global solve & 0.520 (0.0010 / 0.2021) & 0.760 (0.0127 / 0.3087) & 0.761 (0.2558 / 0.3985) & (0.00046 / 0.0029) & 41.1\,s (39.8) & 10.2\,s (9.5) & 1.18\\
 & QCap, half-space & 0.370 (0.00075 / 0.1769) & 0.648 (0.0156 / 0.2702) & 0.612 (0.0532 / 0.3487) & (0.00028 / 0.0034) & as windows & & \\
 & solver-free & 2.271 (0.0033 / 3.2116) & 12.071 (1.3688 / 4.9058) & 11.424 (1.9526 / 6.3312) & (0.0016 / 0.0988) & 0.67\,s$^{\ast}$ & -- & --\\
 & QCap-Fast & 1.549 (0.0023 / 1.6488) & 7.259 (0.3936 / 2.5186) & 4.627 (1.3285 / 3.2490) & (0.0011 / 0.0752) & 0.68\,s$^{\ast}$ & -- & --\\
\hline
35 & QCap, windows & 0.209 (0.00025 / 0.1448) & 0.952 (0.0159 / 0.3185) & 0.975 (0.2420 / 0.4315) & (0.00016 / 0.0031) & 32.7\,s (29.9) & 6.6\,s (5.7) & 3.32\\
 & QCap, global solve & 0.249 (0.00027 / 0.1825) & 1.208 (0.0238 / 0.4013) & 1.609 (0.2969 / 0.5437) & (0.00016 / 0.0031) & 134.0\,s (131.2) & 48.2\,s (47.3) & 1.68\\
 & QCap, half-space & 1.009 (0.0015 / 0.1449) & 0.952 (0.0159 / 0.3185) & 0.975 (0.2420 / 0.4315) & (0.0012 / 0.0053) & as windows & & \\
 & solver-free & 1.948 (0.0022 / 2.7358) & 15.738 (0.4090 / 6.0170) & 13.217 (1.4570 / 8.1511) & (0.0015 / 0.0573) & 2.17\,s$^{\ast}$ & -- & --\\
 & QCap-Fast & 2.471 (0.0026 / 0.7508) & 10.393 (0.6734 / 1.6512) & 6.498 (1.1664 / 2.2322) & (0.0021 / 0.0539) & 2.20\,s$^{\ast}$ & -- & --\\
\hline
67 & QCap, windows & 0.105 (0.00011 / 0.1041) & 0.872 (0.0183 / 0.3053) & 1.186 (0.2940 / 0.4211) & (0.000089 / 0.0022) & 138.6\,s (130.3) & 14.7\,s (12.9) & 3.31\\
 & QCap, global solve & 0.113 (0.00012 / 0.1191) & 1.119 (0.0219 / 0.3494) & 1.294 (0.2802 / 0.4820) & (0.000097 / 0.0025) & 631.6\,s (623.3) & 265.0\,s (263.2) & 3.68\\
 & QCap, half-space & 0.960 (0.0012 / 0.1042) & 0.872 (0.0183 / 0.3053) & 1.186 (0.2940 / 0.4211) & (0.0010 / 0.0046) & as windows & & \\
 & solver-free & 1.304 (0.0013 / 2.8477) & 18.403 (0.4320 / 8.3526) & 16.595 (1.9905 / 11.5230) & (0.0012 / 0.0471) & 5.47\,s$^{\ast}$ & -- & --\\
 & QCap-Fast & 1.408 (0.0014 / 0.9404) & 14.875 (0.5690 / 2.7583) & 6.048 (1.1617 / 3.7966) & (0.0012 / 0.0676) & 5.54\,s$^{\ast}$ & -- & --\\
\end{tabular}
\end{ruledtabular}
\end{table*}

\begin{figure*}[!tbp]
\centering
\includegraphics[width=\textwidth]{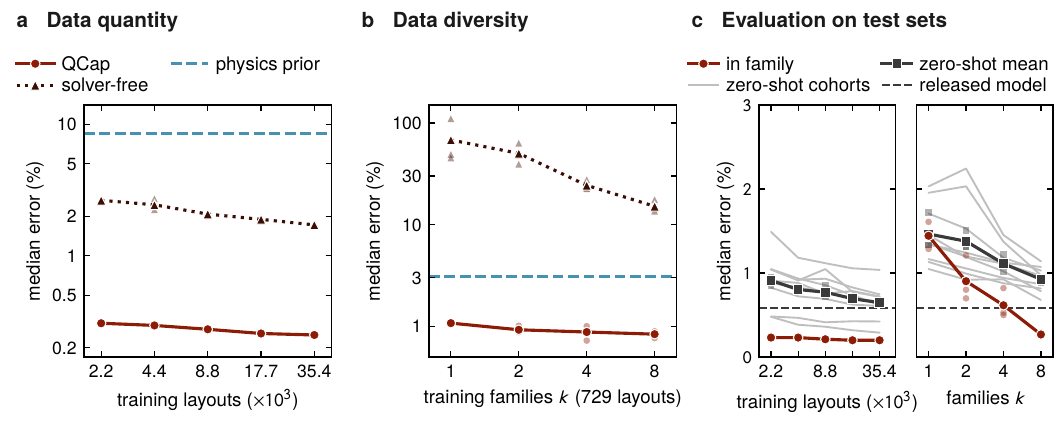}
\caption{How the error of the model scales with the amount and the variety of its training data. (a) Validation error against nested fractions of the 35,354 training layouts (6.25\% to 100\%), given as the number of layouts in thousands. (b) Error on the validation layouts of Xmon, TransmonPocket and shielded concentric, which these runs never train on, against the number $k$ of training families at a fixed 729 layouts. (c) The runs of (a) and (b) on the in-family test set and on the nine zero-shot test sets of Table~\ref{tab:cohorts} (gray, with their mean in black). The dashed line is the zero-shot mean of the released model, a three-seed ensemble trained on all of the data. Every point is the mean of three single networks.}
\label{fig:datacurves}
\end{figure*}

\subsection{Practical workflows, quantities, and outcomes}
\label{app:m-circuit}

For the Xmon, with the cross as conductor S and the claw as conductor N, the charging energy is $E_C=e^2/2hC_{SS}=19.370\,\mathrm{GHz\,fF}/C_{SS}$, where $C_{SS}$ is the cross's diagonal entry, and the claw coupling $\CNS$ is minus the Maxwell entry between claw and cross. For the TransmonPocket, with its two floating pads $s$ and $b$, $E_C=e^2/2hC_\Sigma$ is the charging energy of the differential mode, with
\begin{equation}
C_\Sigma=\frac{C_{ss}C_{bb}-C_{sb}^2}{C_{ss}+C_{bb}+2C_{sb}},
\end{equation}
and the effective coupling of coupler $k$ to that mode is
\begin{equation}
C_{\mathrm{eff},k}=\frac{|c_{ks}C_{b\Sigma}-c_{kb}C_{s\Sigma}|}{C_{s\Sigma}+C_{b\Sigma}},
\label{eq:ceff}
\end{equation}
where $c_{ks}$ and $c_{kb}$ are the Maxwell entries of coupler $k$ with the two pads, $C_{s\Sigma}=C_{ss}+C_{sb}$ and $C_{b\Sigma}=C_{bb}+C_{sb}$.

For the Xmon screen, we solved in Q3D the candidate the model ranked best for each target. For the target $(200\MHz, 3\fF)$ the pick has $E_C=200.35\MHz$ and $\CNS=3.043\fF$, for $(250\MHz, 4\fF)$ it has $246.63\MHz$ and $3.971\fF$, and for $(300\MHz, 2\fF)$ it has $296.75\MHz$ and $1.949\fF$. Every quantity is within 2.6\% of its target. These misses come from how densely the 5000 random candidates cover each target and not from the model, since Q3D agrees with the model's prediction for every pick to within 0.3\% (Sec.~\ref{sec:res-workflows}).

For the TransmonPocket, the initial screen of the candidates took about 25\,min of wall time on 64 CPUs of our cluster and found no candidate within 10\% of all four targets at once. Each refinement round takes the 20 designs the model ranks best, makes 49 variants of each (39 from the third round on), keeps the parents and runs the model on this set again. Four rounds found the candidate of Sec.~\ref{sec:res-workflows}, the last one with five times smaller random changes. The first two rounds took 12\,min on 32 CPUs of our cluster and the last two 21\,min on a laptop with ten processes.

\section{Performance against data quantity and diversity}
\label{app:res-scaling}

We want to know how much of QCap's accuracy comes from the amount of its training data and how much from its variety. To understand this, we train single networks of QCap on subsets of the training data in two studies. In the first we vary the amount, from 6.25\% to all 35354 training layouts in nested fractions that keep the mix of families, and predict on the 4608 validation layouts. In the second we keep the amount fixed at 729 layouts, balanced over the families they are chosen from, and vary the number of families, $k=1$, 2, 4 or 8 of interdigital, paddle, concentric, star, multipad star, single conductor in a ground cut, random shape placements and mixed-shape layouts. These runs never see Xmon, TransmonPocket or shielded concentric and are scored on the 93 validation layouts of those three families. Every level is the mean of three runs, and the solver-free network is trained on the same subsets with the same seeds. QCap-Fast is left out, because its error is set by its learned prior, which is not trained on these subsets. 

Since the 100\% level is a mean of single networks, it is slightly worse than the released three-seed ensemble on the same layouts, 0.250\% (MAE 0.089\fF, RMSE 0.107\fF) against 0.228\% (0.061 / 0.079\fF).

Figure~\ref{fig:datacurves} shows the result of this study. On validation, more data helps QCap a little, lowering its MAE by 0.078\fF\ from 6.25\% to all of the data, and helps the solver-free network much more. More families help both networks, the solver-free one by far the most. Evaluating these models on the in-family test set and on the nine zero-shot test sets of Table~\ref{tab:cohorts} allows us to probe which kind of data makes the model better on layouts it was not trained on (Figure~\ref{fig:datacurves} (c)). More families help far more than more of the same. Going from one family to eight, at the same 729 layouts, lowers the mean error over the nine zero-shot test sets from 1.46\% to 0.92\% (MAE 2.73 to 1.35\fF, RMSE 2.97 to 1.53\fF), while going from 6.25\% to all of the data lowers it only from 0.91\% to 0.65\% (MAE 1.79 to 1.10\fF, RMSE 1.94 to 1.21\fF), a clear gain on five of the nine. Three of the nine test sets, the single conductors, random shapes and mixed shapes, come from families that some of the diversity runs train on, so for those runs they are not fully zero-shot. The in-family test set behaves differently. More data changes little there, since the in-family error has largely saturated already at 6.25\% of the data, while more families help a lot, because with one family most of the in-family test families are missing from training. None of these single networks reaches the released model (dashed line), a three-seed ensemble trained on all of the data.

\bibliography{refs}

\end{document}